\documentclass[prd,aps,preprint,nofootinbib,superscriptaddress]{revtex4}

\pdfoutput=1

\usepackage{graphics}
\usepackage{amsmath}
\usepackage{epsfig}
\usepackage{wasysym}
\usepackage{graphicx}
\usepackage{subfig}

\begin{document}

\title{Kinetic Decoupling and Small-Scale Structure\\ in Effective Theories of Dark Matter}

\author{Jonathan M. Cornell}
\email{jcornell@ucsc.edu} \affiliation{Oskar Klein Centre for Cosmoparticle Physics, Department of Physics, Stockholm University, SE-106 91 Stockholm, Sweden} \affiliation{Department of Physics, University of California, 1156 High St., Santa Cruz, CA 95064, USA} \affiliation{Santa Cruz Institute for Particle Physics, Santa Cruz, CA 95064, USA}

\author{Stefano Profumo}
\email{profumo@ucsc.edu}\affiliation{Department of Physics, University of California, 1156 High St., Santa Cruz, CA 95064, USA}\affiliation{Santa Cruz Institute for Particle Physics, Santa Cruz, CA 95064, USA} 

\author{William Shepherd}
\email{wshepher@ucsc.edu}\affiliation{Department of Physics, University of California, 1156 High St., Santa Cruz, CA 95064, USA}\affiliation{Santa Cruz Institute for Particle Physics, Santa Cruz, CA 95064, USA}

\date{\today}

\begin{abstract}
\noindent The size of the smallest dark matter collapsed structures, or protohalos, is set by the temperature at which dark matter particles fall out of kinetic equilibrium. The process of kinetic decoupling involves elastic scattering of dark matter off of Standard Model particles in the early universe, and the relevant cross section is thus closely related to the cross section for dark matter scattering off of nuclei (direct detection) but also, via crossing symmetries, for dark matter pair production at colliders and for pair-annihilation. In this study, we employ an effective field theoretic approach to calculate constraints on the kinetic decoupling temperature, and thus on the size of the smallest protohalos, from a variety of direct, indirect and collider probes of particle dark matter.
\end{abstract}

\maketitle

\section{Introduction}
In the paradigm of weakly interacting massive particles (WIMPs) as dark matter candidates, the abundance of dark matter observed in the Universe stems from thermal decoupling of the dark matter particles in the early universe. This process involves the pair-annihilation and pair-creation of WIMPs going out of {\em chemical} equilibrium, with the resulting number density freezing out and remaining approximately constant per comoving volume to the present age. WIMP models possess the right range of masses and pair-annihilation/creation cross sections to produce a thermal relic density in the same ballpark as the observed dark matter density, a feat often dubbed the ``WIMP miracle'' \cite{wimp_miracle}. 

After chemical decoupling, WIMPs do not cease to interact with the surrounding thermal bath. It is simply their number density which is no longer affected by particle-number-changing processes. WIMPs ($\chi$) continue to scatter off of (Standard Model) particles  in the thermal bath ($f$), thus remaining in {\em kinetic} equilibrium with the thermal bath, up until the relevant elastic processes ($\chi f\leftrightarrow\chi f$) go out of equilibrium, i.e. the rate for such processes falls below the Hubble expansion rate. At that point, WIMPs completely decouple from the thermal bath, free-streaming and slowing down as the Universe keeps expanding. To a first approximation, this is the age when the first gravitationally collapsed dark matter structures form, with typical size on the same order as an Hubble length at that epoch. WIMP {\em kinetic} decoupling thus sets the small-scale cut-off to the dark matter power spectrum (for a recent review see e.g. Ref.~\cite{review}).

Given a WIMP model, it is thus in principle a well-posed question to ask what the small-scale cutoff to dark matter halos (which we hereafter refer to with the symbol $M_{\rm cut}$) is. The cut-off scale is an important quantity in cosmology: if large enough, it could affect significantly how many ``visible'' small-scale structures, such as dwarf galaxies, form, perhaps being relevant to the question of the ``missing satellite problem'' \cite{Strigari:2007ma} or to other issues associated with small scales in cold dark matter cosmologies \cite{Primack:2009jr}. In principle, the small-scale cutoff sets the size of the most numerous dark matter ``mini-halos'', or protohalos, which might be detectable today either with direct \cite{directhalos} or indirect \cite{indirecthalos} dark matter search experiments. Finally, the cutoff scale is highly relevant to the question of the so-called ``boost factor'', as it literally sets the integration cutoff in the calculation of this factor (in practice, the enhancement to the annihilation rate from a given dark matter halo from sub-structure within the halo).

The calculation of the kinetic decoupling temperature $T_{\rm kd}$, and thus of the small-scale cutoff $M_{\rm cut}$ has been carried out in a variety of model-dependent contexts, including supersymmetry \cite{Hofmann:2001bi, Bringmann:2006mu, Profumo:2006bv}, universal extra-dimensions \cite{Profumo:2006bv, Hooper:2007qk}, and models with Sommerfeld enhancement \cite{vandenAarssen:2012ag,Aarssen:2012fx}. It has become clear that WIMP models accommodate a broad variety of kinetic decoupling temperatures, with resulting cutoff scales ranging from $10^3\ M_\odot$ to much less than $10^{-6}\ M_{\odot}$ even only within the limited framework of the minimal supersymemtric extension of the Standard Model \cite{Profumo:2006bv}, where the symbol $M_\odot$ indicates the mass of the Sun. Particle physics details of the WIMP model affect in a highly model-dependent way the kinetic decoupling, producing a wide array of outcomes, but for many particle physics models there is still a decent correlation between certain experimentally accessible quantities such as the direct detection scattering cross section, as explored in Ref.~\cite{Cornell:2012tb}, and $M_{\rm cut}$. 

A possible model-independent route to evaluating ranges for the expected small-scale cutoff is to consider an effective theory description of interactions between Standard Model and dark matter particles, as pursued, recently, in Ref.~\cite{Gondolo:2012vh, Shoemaker:2013tda}. For example, assuming the dark matter is a spin 1/2 fermion, it is simple to write down the complete set of lowest dimensional operators that mediate such interactions. In turn, by assuming that only one single operator is dominating the relevant dark matter interactions, crossing symmetry allows to draw stringent constraints on the allowed effective energy scale associated with the operator, for example from direct dark matter detection or from collider searches. As a result we can robustly set {\em upper limits} to the size of the small-scale cutoff, for each class of operators, as a function of the relevant operator's effective energy scale. This upper limit is quite significant, as cosmologically relevant effects occur only for sufficiently large such cutoffs.

While rather sophisticated codes now exist to reliably calculate $T_{\rm kd}$ (see e.g. \cite{review}), two potentially important ingredients have been only marginally studied thus far: 

(i) scattering off of quarks only, for example in ``lepto-phobic'' theories with suppressed couplings to leptons (this was first partly addressed in Ref.~\cite{Gondolo:2012vh}), and 

(ii) the role of scattering off of pions, for the same class of theories, for kinetic decoupling temperatures below the QCD confinement phase transition. 

In addition, a third aspect that remains entirely unexplored to date is (iii) the relevance of loop-mediated scattering off of leptons, again notably in leptophobic theories.

In the present study, in addition to the general program of setting upper limits to the small-scale cutoff in the context of the mentioned effective theory description of interactions between Standard Model and dark matter particles, we address in detail the three novel issues listed above. We show that for leptophobic theories there exists an interesting interplay between loop-mediated scattering off of leptons and scattering off of pions, and that the two effects are generically comparable. We find that for WIMP models that can be described to a good approximation by an effective operator belonging to the class we consider here, there are stringent upper limits on the cut-off scale to the matter power spectrum, typically on the order of $10^{-3}\ M_\odot$. This scale hints at the fact that WIMP effective theories are not likely to have any impact on small-scale structure issues in cold dark matter cosmology. On the other hand, since the predicted protohalos are typically very small, sizable boost factors from substructure enhancements are a rather generic prediction of effective theories of dark matter.

The reminder of this paper is organized as follows: we outline the class of effective operators we consider in the following section \ref{sec:operators}; we then discuss how we calculate the kinetic decoupling and how we estimate the size of the small-scale halo size cutoff in section \ref{sec:protohalos}; section \ref{sec:results} presents all of our results; and the final section \ref{sec:concl} summarizes our findings and concludes.

\section{Classification of effective operators} \label{sec:operators}

The effective operator framework has been explored as a method for comparing experimental bounds coming from various types of experiments on dark matter couplings to Standard Model fields \cite{EffDM,Fox:2011fx}. Within this framework, one writes down higher-dimensional operators which couple dark matter to quarks, leptons, or Standard Model bosons, requiring that (i) the operator contain at least two dark matter particles to ensure stability, and that (ii) Standard Model gauge symmetries are respected. One operator from the list of possible operators is then assumed to be the dominant one for the physics being investigated, and its effects are explored assuming the other operators are suppressed and, thus, do not contribute to the observables in question. Each operator of interest is investigated separately in this way, and any interference effects from having multiple operators active simultaneously are assumed to be small. Generally these interference effects are equivalent to changing the assumed chirality structure of the operator in question, e.g. interfering a vector and an axial operator with equal suppression scales is equivalent to considering an operator which only couples to one chirality.

The basic assumption of this parametrization of dark matter interactions is that dark matter is the only new field light enough to be kinematically relevant, and these operators are suppressed by a mass scale which is related to the expected mass of the additional particles which mediate the interactions in some more complete model underlying the effective theory. Within the region of parameter space where this assumption is valid, a given complete model can be mapped into the space of these operators by integrating out the additional heavy fields. This assumption is a fairly weak one for elastic scattering of dark matter off of Standard Model particles, where the momentum exchange is typically on the order of the MeV, but is a fairly strong assumption for LHC searches, where the center of mass energy of the created dark matter pair can be quite large compared to the dark matter mass. We therefore encourage caution when considering the collider bounds on these operators, but expect that the results for kinetic decoupling and the bounds arising from direct detection should be robust.

We also calculate the thermal relic density of WIMPs under the assumption that the same operator dominates dark matter interactions with Standard Model particles in the early as well as in the late universe. Of course, this is a rather strong assumption, as it entails for example the absence of processes such as coannihilation, the presence of thresholds or resonances that could exist at finite temperature but not in the late universe, and the absence of temperature-suppressed operators that might dominate the chemical freeze-out while being irrelevant at the later kinetic freeze-out. We note, however, that this assumption is largely equivalent to other assumptions discussed above, where it is presumed that dark matter is the only kinematically relevant new particle in the theory and that one operator is dominant in all of the observables being searched for.

We consider here a subset of all possible operators which conserve parity in addition to the Standard Model gauge symmetries. The operators of interest are
\begin{eqnarray}
\cal{O}_S&=&\frac{m_f}{\Lambda_S^3}\bar\chi\chi\bar ff \label{eq:scalarop}\\
\cal{O}_P&=&\frac{m_f}{\Lambda_P^3}\bar\chi\gamma^5\chi\bar f\gamma^5f \label{eq:pscalarop}\\
\cal{O}_V&=&\frac{1}{\Lambda_V^2}\bar\chi\gamma^\mu\chi\bar f\gamma_\mu f \label{eq:vectorop}\\
\cal{O}_A&=&\frac{1}{\Lambda_V^2}\bar\chi\gamma^\mu\gamma^5\chi\bar f\gamma_\mu\gamma^5f \label{eq:pvectorop} \\
\cal{O}_T&=&\frac{m_f}{\Lambda_T^3}\bar\chi\sigma^{\mu\nu}\chi\bar f\sigma_{\mu\nu}f \label{eq:tensorop},
\end{eqnarray}
where $\Lambda_I$ is the suppression scale for operator $\cal{O}_I$. Note that the operators which are chirality-violating are assumed to be proportional to the fermion mass to preserve $SU(2)_L$ and avoid inducing large effects in low-energy flavor observables. The first four operator normalizations are standard within the effective dark matter literature, but previous searches for contact operators have not included the mass suppression for the tensor operator to better make contact with direct detection bounds. We choose to consider the theoretically better motivated normalization of the tensor operator which does include a quark mass suppression, as the operator is chirality-violating and thus would require an insertion of the Higgs field to respect the SM gauge symmetries. Previous analyses have considered the operator without a quark mass dependence to make better contact with direct detection searches, as the unsuppressed tensor induces a coupling to the spin of the quarks composing the nucleon minus the spin of the antiquarks in the nucleon, but it is not clear how a tensor operator with that normalization would be alligned with the mass basis of the quarks so well as to avoid inducing unacceptably large corrections to flavor observables. The choice to include the quark mass suppression of the tensor operator leaves us without collider and direct detection bounds to compare to, and therefore we will only plot the early universe curves for these operators.

For each operator, we specify which Standard Model fermions the dark matter particle couples to. Generically, leptons are the most significant contributors to keeping the dark matter in kinetic equilibrium with the Standard Model thermal bath, while many of the key experimental searches constrain primarily the couplings to quarks. We choose here to consider explicitly three cases, wherein the dark matter couples only to leptons, only to quarks, or to both with equal suppression scales. For cases including quark couplings we plot the strongest available experimental bounds from LHC searches and direct detection searches, and in cases including lepton couplings we will additionally plot LEP search bounds. In the special case of the lepton-only vector operator we will in addition plot the direct detection bounds induced at one-loop order, as discussed in Ref.~\cite{ZupanKoppetal,Fox:2011fx}.

\section{The formation of protohalos}\label{sec:protohalos}

\subsection{Temperature of kinetic decoupling}\label{sec:tkd}

To calculate the temperature of kinetic decoupling, we use the numerical routine described in Ref.~\cite{review}, which has been integrated into the DarkSUSY code \cite{darksusy}. An effective WIMP temperature parameter is defined in the following form:
\begin{equation} \label{eq:tchi}
T_\chi \equiv \frac{2}{3} \left< \frac{{\bf p}^2}{2 m_\chi} \right> = \frac{1}{3 m_\chi n_\chi} \int \frac{{\rm d}^3 p}{(2 \pi)^3} {\bf p}^2 f({\bf p}).
\end{equation}
In the equation above $m_\chi$ is the WIMP mass and $n_\chi$ is its number density. To determine the time evolution of this parameter, we consider the Boltzmann equation for a flat Friedmann-Robertson-Walker metric:
\begin{equation}
\label{eq:Boltzmann}
E(\partial_t - H {\bf p} \cdot \nabla_{\bf p}) f = C[f].
\end{equation}
Here $f$ is the WIMP phase space density, $E$ and {\bf p} are the comoving energy and 3-momentum respecitively, and $H$ is the Hubble parameter. $C[f]$ is the collision term for a scattering process between a non-relativistic WIMP and a relativistic Standard Model scattering partner. This was shown in Ref.~\cite{review} to be of the form
\begin{equation}
C[f] = c(T) m_\chi^2 \left[m_\chi T \nabla^2_{\bf p} + {\bf p} \cdot \nabla_{\bf p} + 3 \right] f ({\bf p}),
\end{equation}
where
\begin{equation}
\label{cTdef}
c(T) =  \sum_i\frac{g_\mathrm{SM}}{6(2\pi)^3m_\chi^4T} \int dk\,k^5 \omega^{-1}\,g^\pm\left(1\mp g^\pm\right)\mathop{\hspace{-11ex}\overline{\left|\mathcal{M}\right|}^2_{t=0}}_{\hspace{4ex}s=m_\chi^2+2m_\chi\omega+m_\ell^2}\,.
\end{equation} 
In Equation (\ref{cTdef}), the sum is taken over all possible Standard Model scattering partners, $g_{\rm SM}$ is the number of associated spin degrees of freedom, $\omega$ is the energy of the Standard Model particle and $k$ its momentum, and $g^\pm$ is the distribution for Fermi or Bose statistics, $g^\pm(\omega) = (e^{\omega/T} \pm 1)^{-1}$. In all expressions above, the upper sign is for fermions and the lower is for bosons. $\overline{\left|\mathcal{M}\right|}^2$ represents the scattering amplitude squared, summed over final and averaged over initial spin states. Detailed calculations of $\overline{\left|\mathcal{M}\right|}^2$ for all relevant cases for our results are included in the appendices.

As kinetic decoupling can take place either before or after the QCD phase transition at $T_c \approx 170 ~{\rm MeV}$, we need to consider carefully the effects of quark confinement on the above sum. At temperatures before $4 \, T_c$, we follow the convention of Ref.~\cite{review}, where the WIMPs scatter off leptons and, to be conservative, the three lightest quarks. After $4 \, T_c$, we no longer consider scattering off quarks. We however extend the treatment of Ref.~\cite{review} by including scattering of the dark matter off pions after the QCD phase transitions for the cases in which this process occurs at leading order, i.e. for the scalar, Eq.~(\ref{eq:scalarop}), and vector, Eq.~(\ref{eq:vectorop}), operator cases. Also, it is important to note in the above expression for $c(T)$, the scattering amplitude is evaluated in the $t= 0$ limit, where $t$ is the squared difference between the incoming and outgoing 4-momenta of a scattering particle. This limit is reasonable because the average momentum transfer in a scattering event between a relativistic particle and a heavy WIMP should be quite small. However, for the pseudoscalar case, Eq.~(\ref{eq:pscalarop}), the scattering amplitude vanishes for forward scattering, so we need to consider the scattering amplitude when the momentum transfer is not zero. Ref.~\cite{Gondolo:2012vh} introduced a method to average over all possible values of $t$, in which $c(T)$ now takes the form:
\begin{equation}
c(T) =  \sum_i\frac{g_\mathrm{SM}}{6(2\pi)^3m_\chi^4T} \int dk\, k^5 \omega^{-1}\,g^\pm\left(1\mp g^\pm\right) \frac{1}{\left(4 k^2 \right)^2} \int_{-4k^2}^0 dt (-t) \mathop{\overline{\left|\mathcal{M}\right|}^2_{\hspace{0ex}s=m_\chi^2+2m_\chi\omega+m_\ell^2}}.
\end{equation}

Returning now to $T_\chi$ in Eq.~(\ref{eq:tchi}), to find the differential equation which describes its evolution, we multiply Eq.~\ref{eq:Boltzmann} by ${\bf p}^2/E$ and integrate over ${\bf p}$ to find
\begin{equation} \label{eq:diff}
\left( \partial_t + 5 H \right) T_\chi = 2 m_\chi c(T) \left(T-T_\chi \right).
\end{equation}
The behavior of $T_\chi$ has two limiting cases: at high temperatures when $T_\chi = T$ and at low temperatures when $T_\chi$ changes only because of the expansion of the universe, i.e. $T_\chi \propto a^{-2}$, and the kinetic decoupling temperature is when there is a rapid change between these two behaviors. As described in Ref.~\cite{review}, a code has been developed to numerically integrate Eq.~(\ref{eq:diff}) and find this transition temperature, and we use this routine to calculate $T_{\rm kd}$.

\subsection{Protohalo Size}

There are two mechanisms which independently set a limit on the smallest possible protohalo mass, $M_{\rm cut}$: (i) the free streaming of WIMPs after kinetic decoupling and (ii) the coupling of the WIMP fluid to acoustic oscillations in the SM particle heat bath. In determining our limit on the protohalo mass, we use the outcome of these two processes giving the largest (hence dominant) ${M_{\rm cut}}$.

\subsubsection{Viscosity and Free Streaming}
At kinetic decoupling, the decoupling of the WIMP fluid from the SM particle fluid leads to viscosity between the two fluids that cause density perturbations in the WIMP fluid to be damped out \cite{Hofmann:2001bi}. After $T_{\rm kd}$, the WIMPs free stream from areas of high to low density, causing further damping of the perturbations. The net result of these processes is an exponential damping of the perturbations with a characteristic comoving wavenumber \cite{green,Green:2005fa}:
\begin{equation}
\label{eq:fs}
k_\mathrm{fs} \approx \left( \dfrac{m}{T_\mathrm{kd}} \right)^{1/2} \dfrac{a_\mathrm{eq} / a_\mathrm{kd}}{\ln (4 a_\mathrm{eq} / a_\mathrm{kd})} \dfrac{a_\mathrm{eq}}{a_\mathrm{0}} H_\mathrm{eq}.
\end{equation}
In the Equation above the ``eq'' subscript signifies that the quantity should be evaluated at matter-radiation equality. To find the resulting mass cutoff from these effects, one just calculates the mass contained in a sphere of radius $\pi/k_\mathrm{fs}$, i.e. \cite{review}:
\begin{equation}
M_\mathrm{fs} \approx \dfrac{4 \pi}{3} \rho_\chi \left( \dfrac{\pi}{k_\mathrm{fs}} \right)^3 = 2.9 \times 10^{-6} M_{\odot} \left( \frac{1 + \ln \left(g_\mathrm{eff}^{1/4} T_\mathrm{kd}/ \mathrm{30 \ MeV} \right)/18.56} {\left(m_\chi / \mathrm{100 \ GeV} \right)^{1/2} g_\mathrm{eff}^{1/4} \left(T_\mathrm{kd}/\mathrm{50 \ MeV} \right)^{1/2}} \right)^3.
\end{equation}
In the above equation $g_{\rm eff}$ is the number of effective degrees of freedom in the early universe evaluated at $T_{\rm kd}$.

\subsubsection{Acoustic Oscillations}
It has also been noted that the density perturbations in the WIMP fluid, coupled to the SM particle fluid before $T_{\rm kd}$, should oscillate with the acoustic oscillations in the heat bath. At kinetic decoupling, modes of oscillation with $k$ values large enough that they have entered the horizon are damped out, while modes with $k$ values corresponding to scales larger than the horizon size grow logarithmically \cite{loebzalda,Bertschinger:2006nq}. Therefore, the characteristic damping scale is just the size of the horizon at kinetic decoupling ($k_{\rm ao} \approx \pi H_{\rm kd}$), and there is another cutoff mass from this process of the form \cite{review}:
\begin{equation}
\label{eq:mao}
M_\mathrm{ao} \approx \dfrac{4 \pi}{3} \dfrac{\rho_\chi}{H^3} \bigg|_{T=T_\mathrm{kd}} = 3.4 \times 10^{-6} M_\odot \left( \dfrac{T_\mathrm{kd} g_\mathrm{eff}^{1/4}}{50 \; \mathrm{MeV}} \right)^{-3}.
\end{equation}

\section{Results}\label{sec:results}
As discussed above, for each effective operator in Eq.~(\ref{eq:scalarop}-\ref{eq:tensorop}) we consider three cases as far as the relevant Standard Model particle class the dark matter couples to:
\begin{enumerate} 
\item Universal couplings to all SM fermions;
\item Couplings to leptons only;
\item Couplings to quarks only. 
\end{enumerate}
Each case presents distinct behaviors in the early as well as in the late universe, and leads to different constraints and conclusions for the effective cutoff scale. Leptonic couplings, when present, tend to dominate the process of kinetic decoupling, as a simple result of the fact that at the relevant temperatures leptons (especially electron/positron and neutrinos) are in a relativistic state and the number densities are not Boltzmann-suppressed. On the other hand, quark couplings lead to stronger bounds from colliders and direct detection. If lepton couplings are absent then the contributions of quark couplings to kinetic decoupling must be treated with care due to the QCD confinement phase transition. Before the phase transition there is a thermal bath of quarks and the calculation of the scattering rate proceeds analogously with that for the leptonic couplings, but after the phase transition pions are the dominant hadrons and the matrix element of the quark bilinear in the pion must be evaluated. In addition, loop-induced scattering off of leptons arises generically even for vanishing direct couplings to leptons. This effect, which has never been considered in this context before, competes with scattering off of pions, and becomes more and more relevant as pions become less and less abundant at decreasing temperatures due to Boltzmann suppression. We will discuss each operator's coupling to pions individually in presenting our results.

For each case we also present all relevant bounds on effective dark matter interactions from collider searches both at the LHC \cite{atlasconf, cmspas} and at LEP \cite{Fox:2011fx}. These constraints are subject to the concerns described in section \ref{sec:operators} regarding the possibility of probing additional particles at colliders due to the large center of mass energies involved. For all operators which lead to appreciable direct detection cross sections we also plot the current leading bounds from those experiments. For spin-independent scattering the current leading bounds come from the Xenon 100 experiment \cite{Aprile:2012nq}, while for spin-dependent scattering they are set by the SIMPLE \cite{Felizardo:2011uw} and PICASSO \cite{Archambault:2012pm} experiments.

For all plots we also present relic density constraints. The line on the plots corresponds to when $\Omega_\chi h^2 = 0.1189$, the best fit value quoted by the Planck collboration \cite{planck} when combining their CMB results, WMAP polarization results, high-$\ell$ CMB data from ground telescopes and baryon acoustic oscillation measurements.
For all operators except the tensor case (which has no simple, tree-level UV completion) we use the micrOMEGAs code \cite{Belanger:2010pz,Belanger:2006is} to calculate the relic density. This was checked analytically to correspond with setting the annihilation cross section to the appropriate value $\langle\sigma v\rangle\approx3\times10^{-26}~{\rm cm^3/s}$, and this analytical requirement was used to calculate the relic density requirement for the tensor operator case.

\subsection{Scalar Operator}

\captionsetup[figure]{position=t}
\captionsetup[subfigure]{position=b}
\begin{figure}
\caption{\label{fig:Scalar} \it \small Plots of contours of constant $T_{\rm kd}$ (left) and $M_{\rm cut}$ (right) for the case of a scalar operator interaction between WIMPs and SM particles. Shaded regions and the dashed curve represent regions of parameter space excluded by collider and direct detection results, while the solid gray curve represents the correct dark matter relic density.}
		\subfloat [\it \small DM scatters off all SM fermions before the QCD phase transition and leptons and pions after.]{
			\includegraphics[width=.49\textwidth]{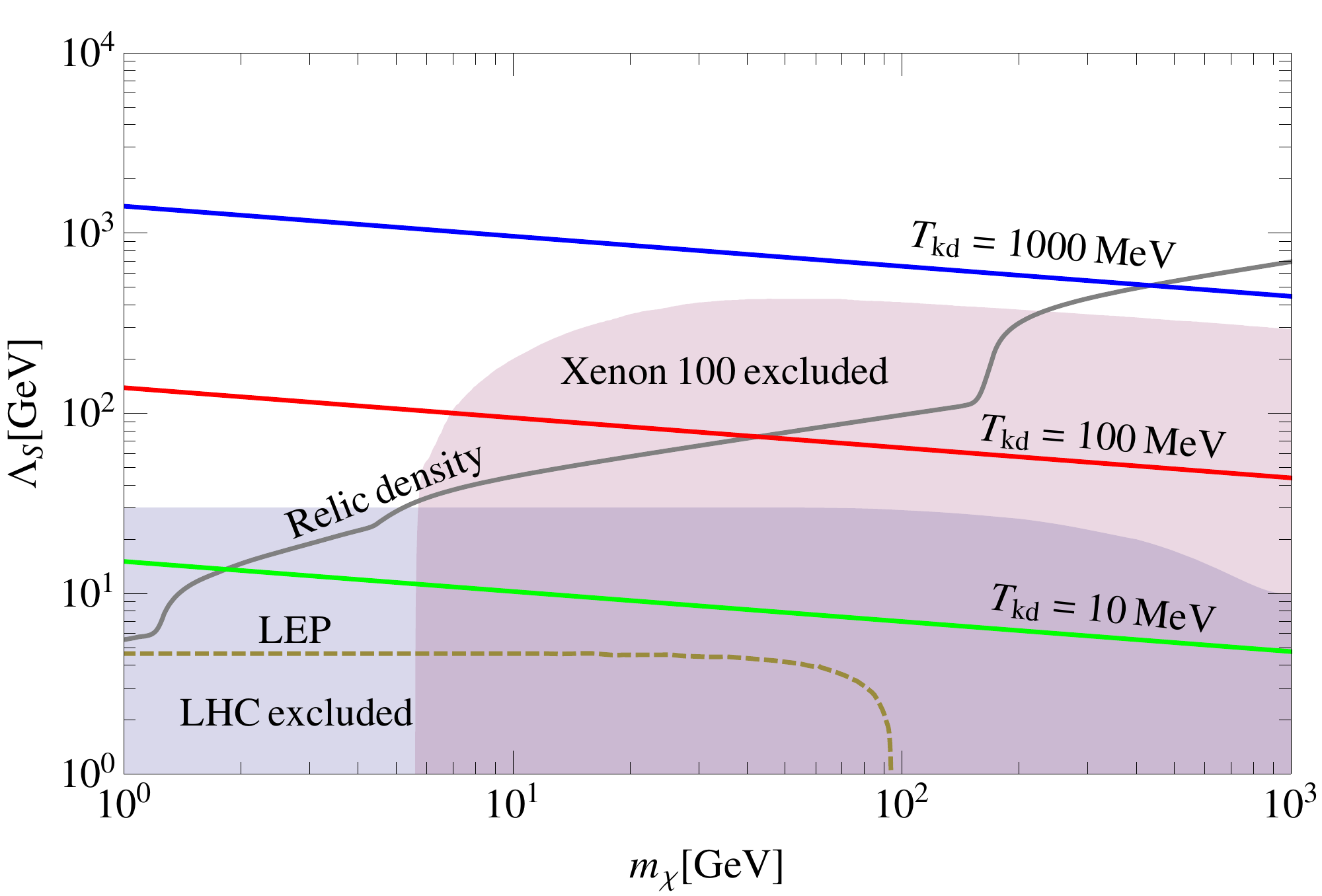}
			\includegraphics[width=.49\textwidth,clip]{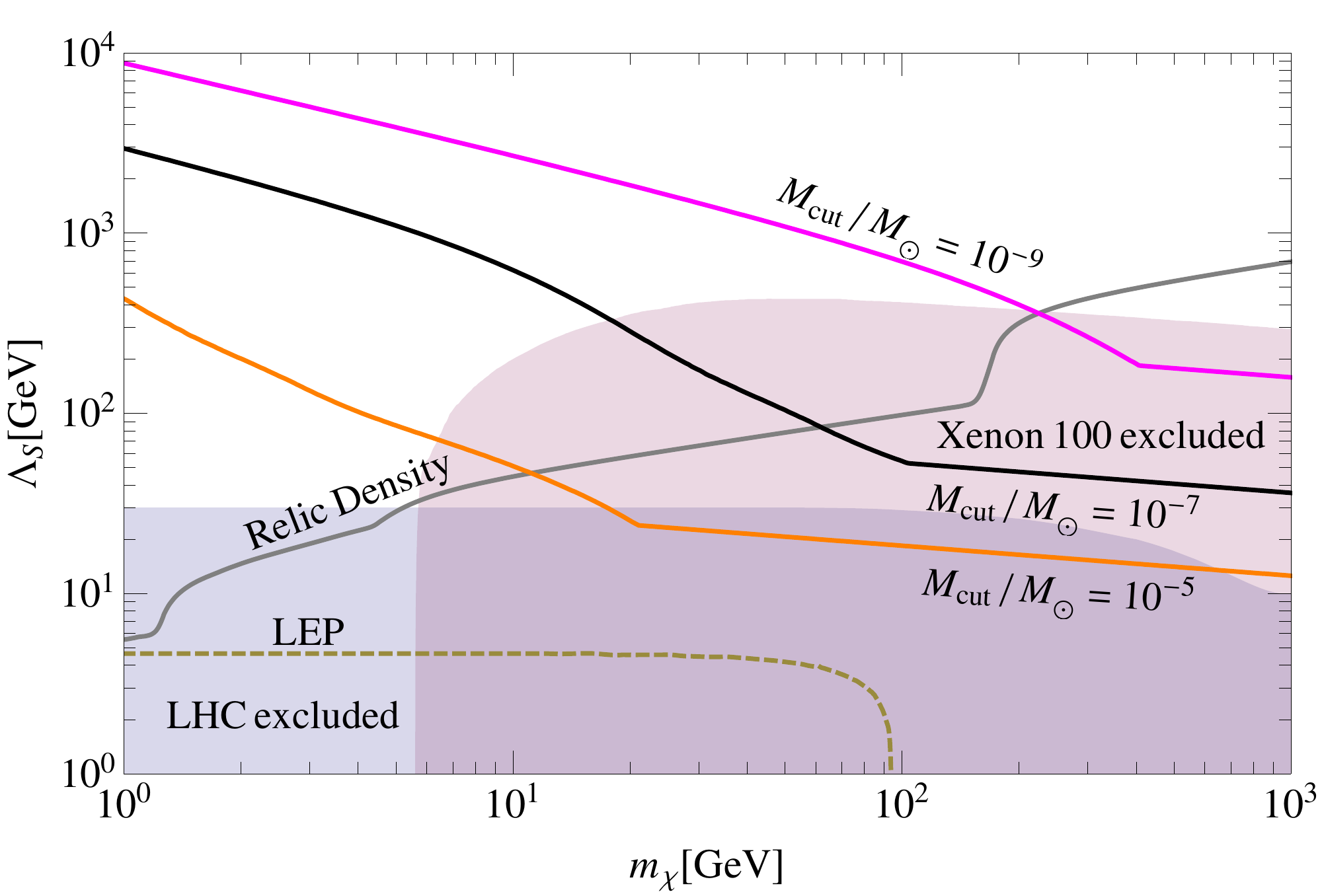}
			\label{fig:Scalar_univ}}\hfill
		\subfloat [\it \small DM scatters off leptons only.]{
			\includegraphics[width=.49\textwidth,clip]{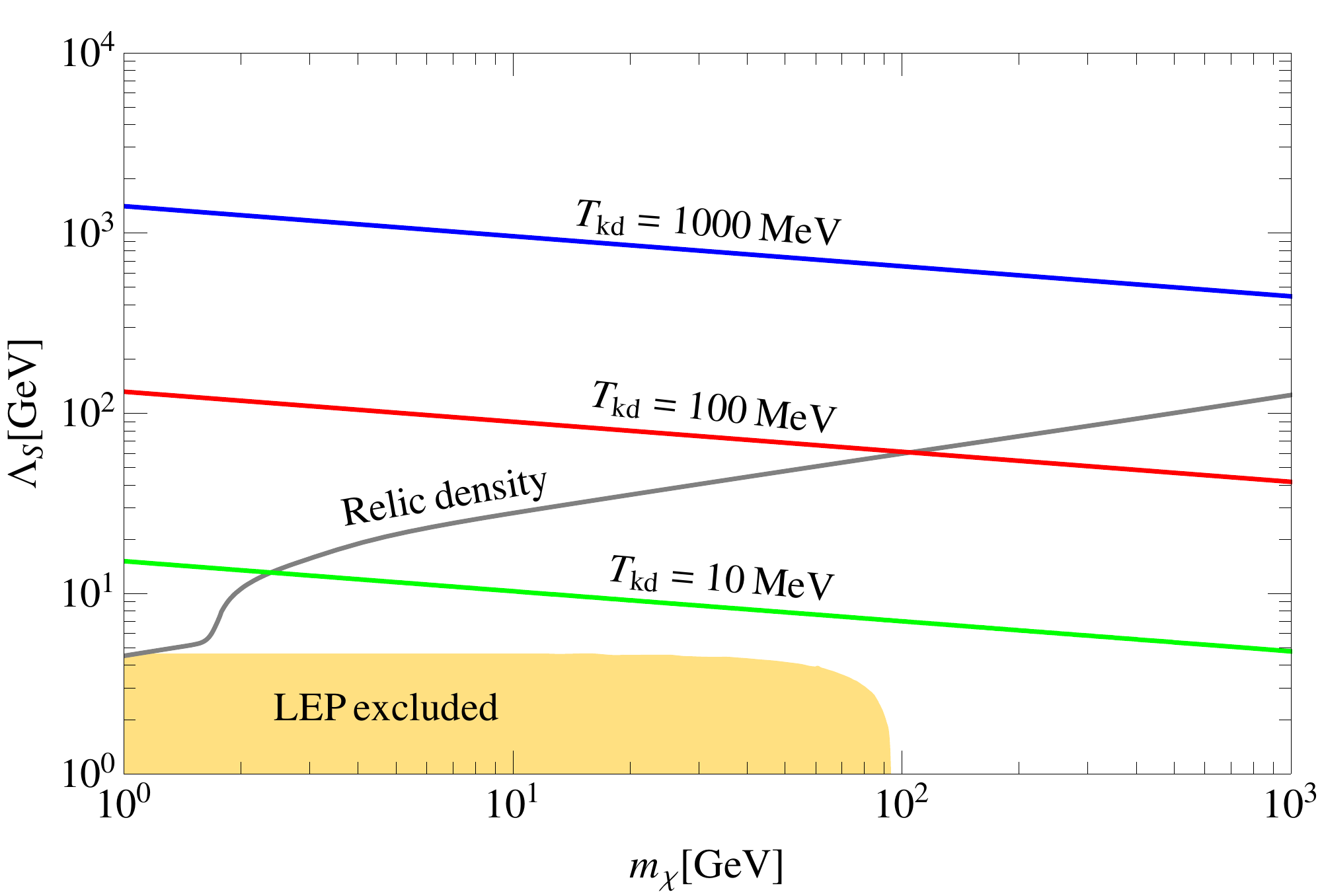}
			\includegraphics[width=.49\textwidth,clip]{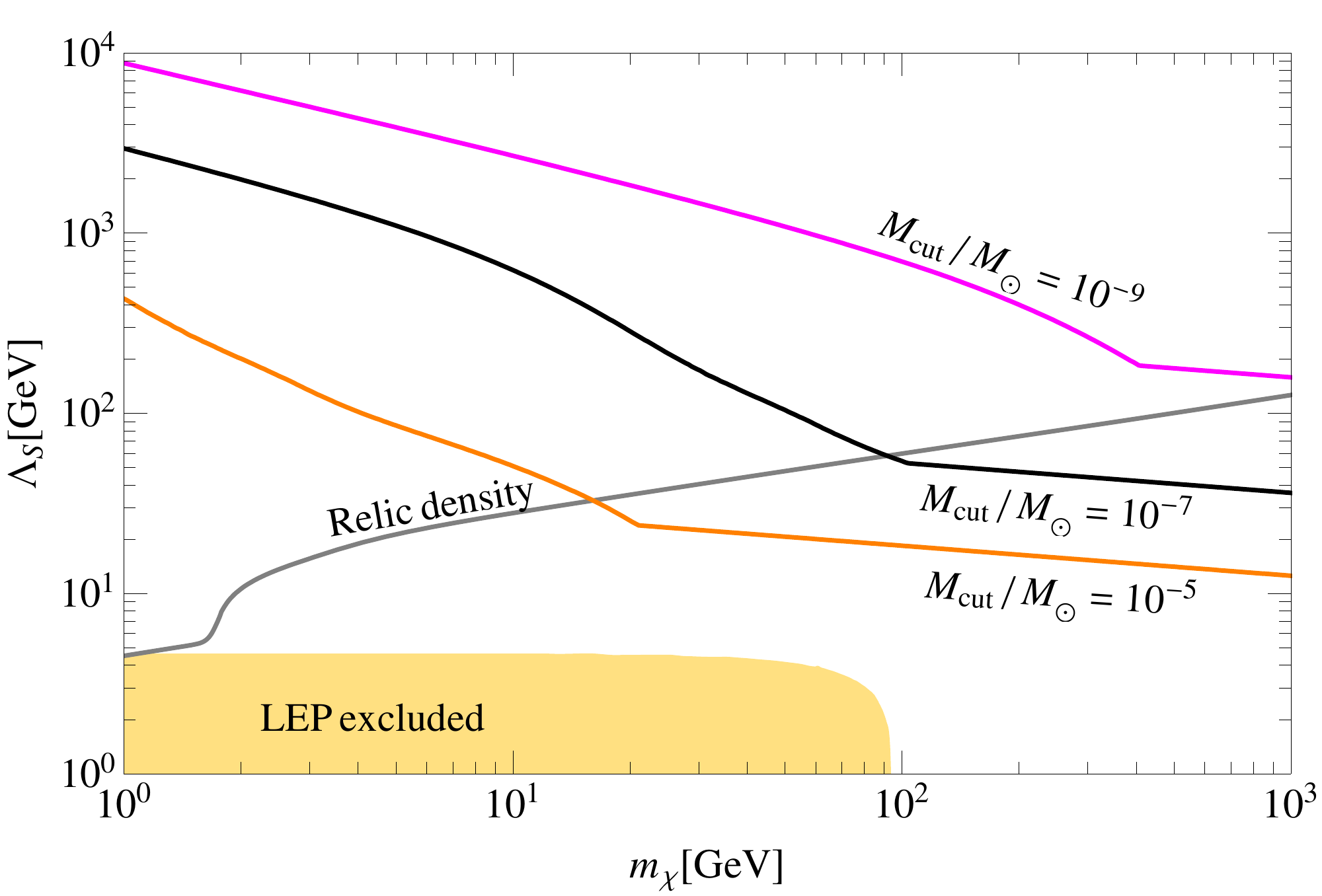}
			\label{fig:Scalar_leptons}}\hfill
		\subfloat [\it \small DM scatters off quarks before the QCD phase transition and pions after.\label{fig:Scalar_quarks}]{
			\includegraphics[width=.49\textwidth,clip]{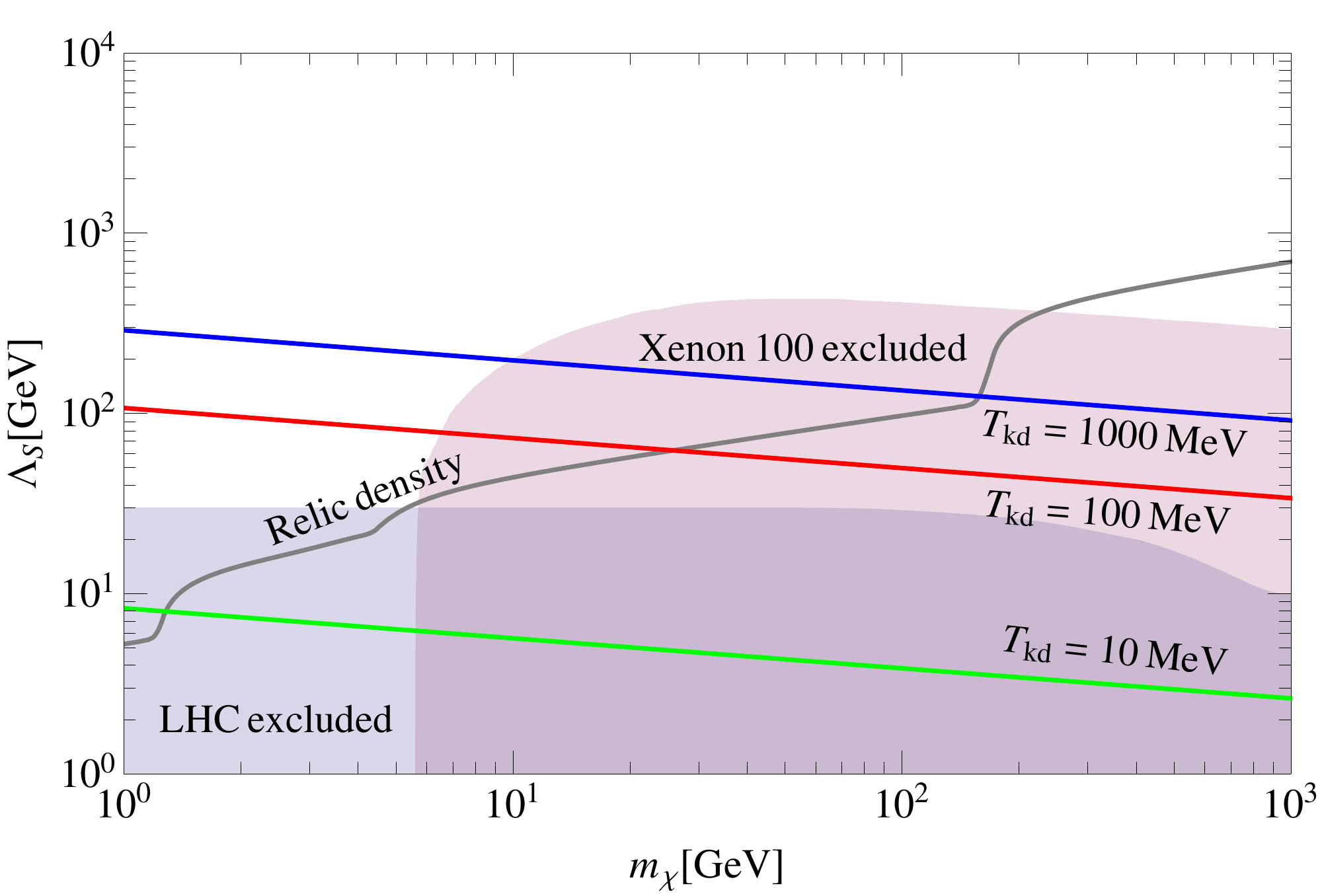}
			\includegraphics[width=.49\textwidth,clip]{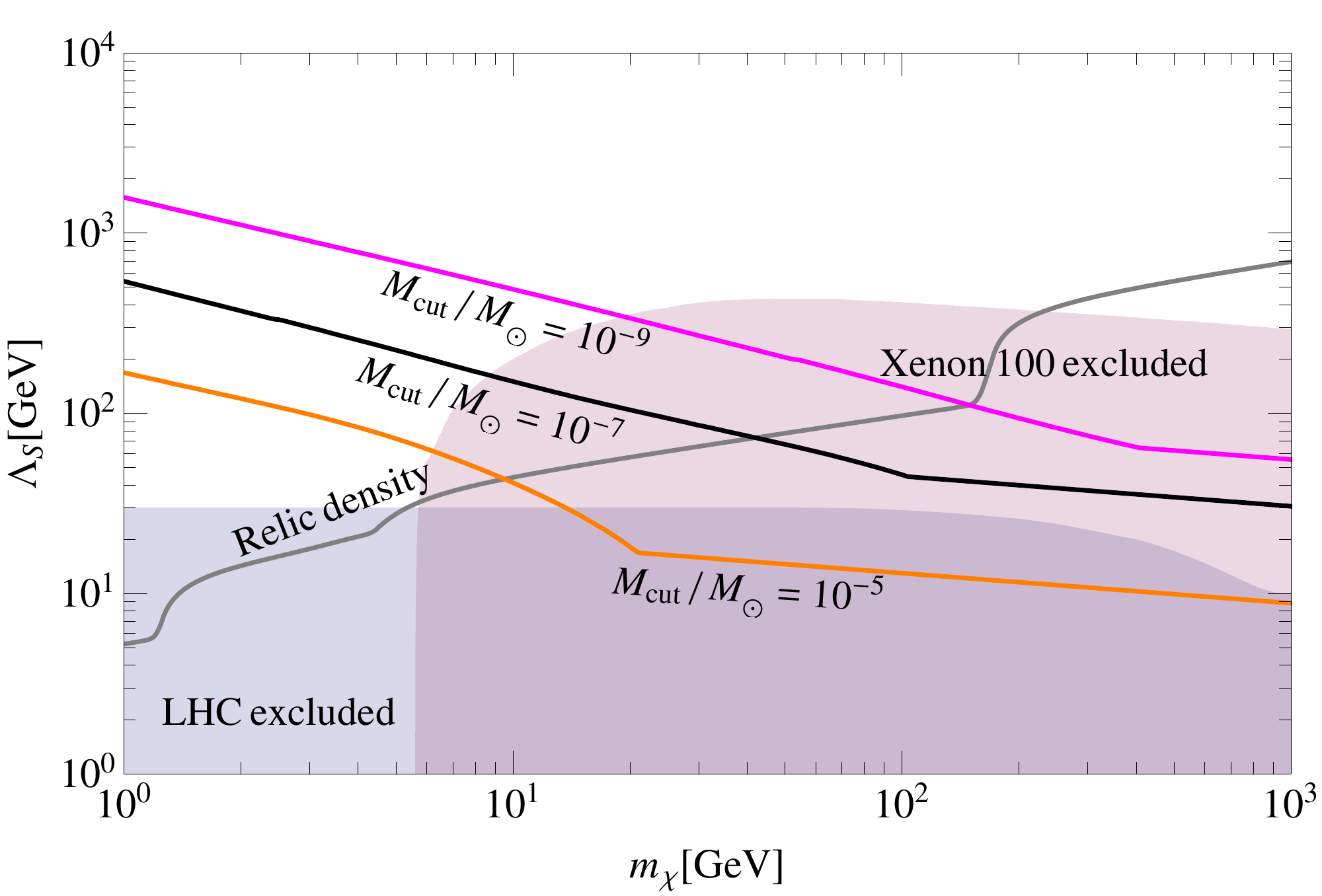}}\hfill
\end{figure}

The scalar-type coupling of dark matter to SM fermions contributes to direct detection in the case of quark couplings, and has been constrained by collider searches for both quark and lepton couplings. The collider constraints are relatively weak in this case, however, because of the mass-suppression of this chirality-violating operator. While pair annihilation, direct detection, and scattering responsible for kinetic decoupling all have access to the heavier SM fermion generations, the initial state, for collider studies, is dominated by the lighter states, and therefore collider bounds are weakened relative to the other dark matter probes.

For the case of universal coupling to SM fermions through the scalar operator the results are presented in figure \ref{fig:Scalar_univ}. We note that for dark matter above approximately 10 GeV in mass the bounds from direct detection, which are strongest in that region, indicate that the kinetic decoupling temperature must be on the order of 1 GeV. The resulting cutoff scale for the smallest protohalos is on the order of the Earth mass (about $10^{-6}\ M_{\odot}$) for WIMP masses above 10 GeV. The relic density matches the observed dark matter only for masses above 200 GeV. Models that possess the right thermal relic density have extremely suppressed cutoff scales, smaller than $10^{-9}\ M_\odot$ (see the right panel of figure \ref{fig:Scalar_univ}).

For the case of lepton-only couplings, the only relevant bound on this operator is from LEP, and the resulting bound is weak enough to not significantly constrain the process of kinetic decoupling. The corresponding results are shown in figure \ref{fig:Scalar_leptons}, which indicates that kinetic decoupling temperatures below 10 MeV are possible in this case, resulting in small-scale cutoffs exceeding the Earth mass. We estimate in this case that the largest possible cutoff mass scale is of about $M_{\rm cut}\sim 10^{-3}\ M_\odot$. We also note that the LEP limits do not impact the cutoff scales $\Lambda_S$ needed to produce the correct thermal relic density. For WIMP masses of about 10 GeV we find that models that have the correct thermal relic density produce a small scale cutoff of $10^{-5}\ M_\odot$, while those with a mass of 100 GeV of about $10^{-7}\ M_\odot$ and those with a mass of 1 TeV of approximately $10^{-9}\ M_\odot$.

Finally, for quark-only couplings the matrix element $\langle\pi|\bar qq|\pi\rangle$, implicitly summed over quark flavors, is needed to evaluate the coupling to pions after the QCD phase transition. This has been evaluated previously in the context of contributions to direct detection by \cite{kamionkowskietal} using soft-pion techniques to be
\begin{equation}
\langle\pi^a|\bar qq|\pi^a\rangle=\frac{m_\pi^2}{2}\langle\pi^a|\vec\pi\cdot\vec\pi|\pi^a\rangle,
\end{equation}
where $\vec\pi$ is a pion iso-vector, Eq.~(\ref{eq:pionvec}). We have implemented this scattering amplitude for interactions after the QCD phase transition with pions, which are the dominant components of the thermal bath at the relevant temperatures. The results for quark-only couplings are shown in figure \ref{fig:Scalar_quarks}. Direct detection forces in this case the size of the smallest protohalos to values well below $10^{-9}\ M_{\odot}$ for dark matter particle masses larger than 20 GeV. Models with the correct relic density must have masses above 200 GeV, and small scale cutoff smaller than $10^{-11}\ M_{\odot}$ in this case.

\subsection {Vector Operator}

\begin{figure*}[!t]
\begin{minipage}{.49\linewidth}
\includegraphics[width=1.0\textwidth,clip]{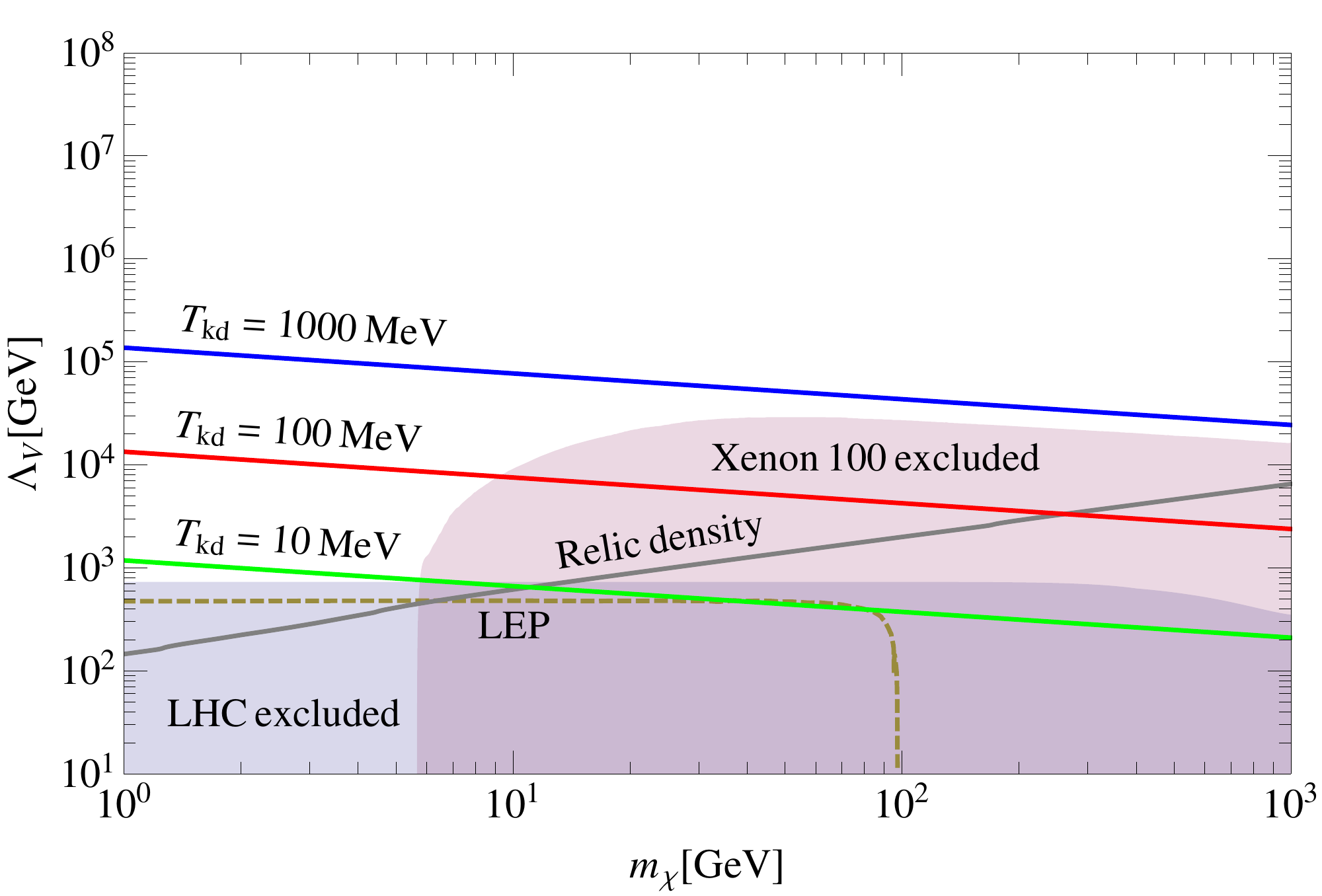}
\end{minipage}
\begin{minipage}{.49\linewidth}
\includegraphics[width=1.0\textwidth,clip]{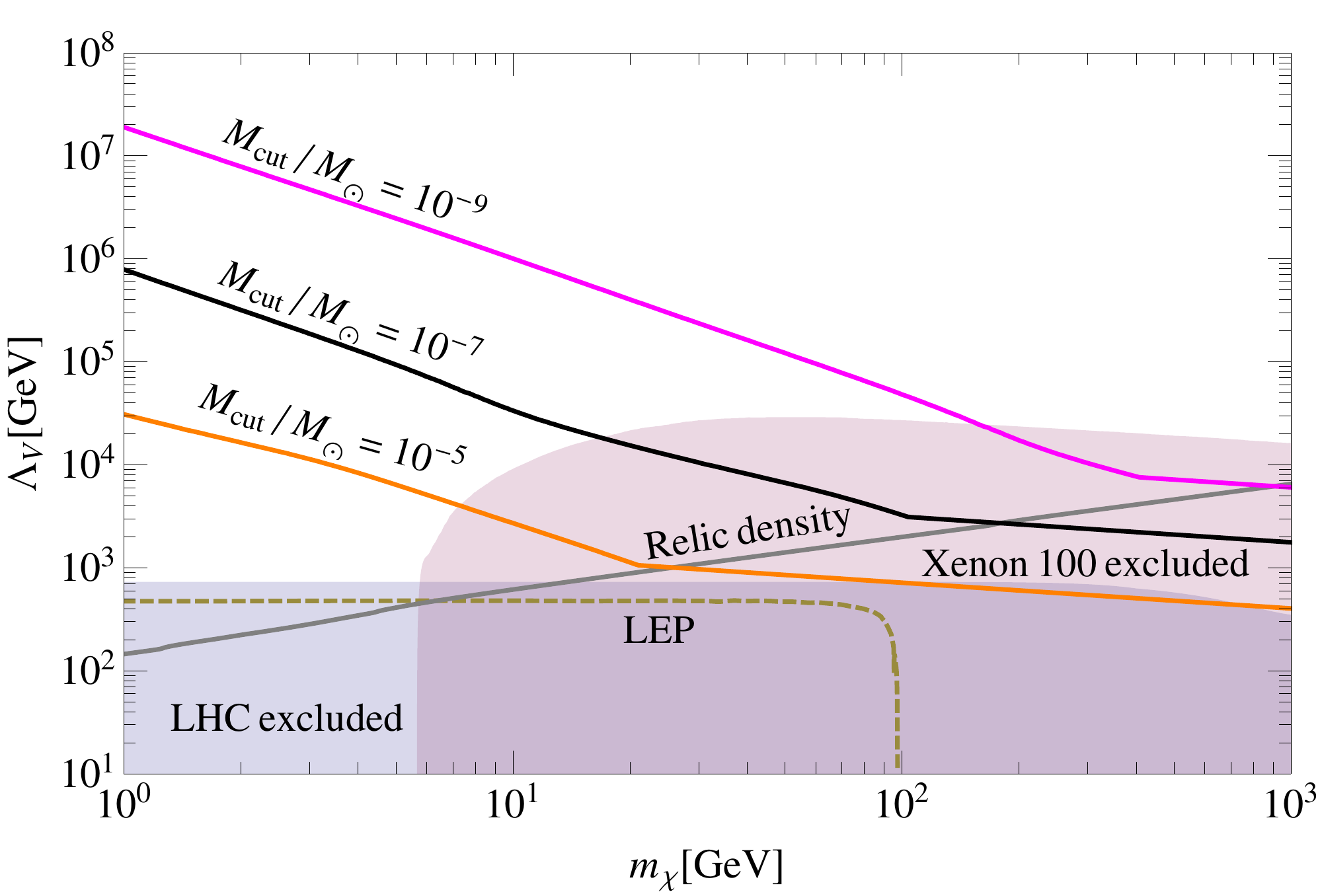}
\end{minipage}
\caption{\label{fig:Vector_univ}\it\small Plots of contours of constant $T_{\rm kd}$ (left) and $M_{\rm cut}$ (right) for the case of a vector operator interaction between WIMPs and SM fermions in which WIMPs couple with the same strength to all SM fermions. Shaded areas signify the regions of parameter space excluded by LHC and direct detection results, and the dashed line corresponds to a limit on the parameters from an analysis of LEP data. The solid gray curve represents the correct relic density.}
\end{figure*}

Vector operators are generically better constrained by collider searches than scalar operators are, but are also very tightly constrained by direct dark matter detection. For universal couplings, direct detection is again the dominant constraint for dark matter masses above about 10 GeV, and those constraints again force us to conclude that the kinetic decoupling temperature must be of order 1 GeV. We show our results in figure \ref{fig:Vector_univ}. We do not find any sub-TeV WIMP models with a viable thermal relic density, if this operator is the only important contributor to chemical freeze-out. For masses above 100 GeV, we find that the cutoff scale is always smaller than approximately $10^{-9}\ M_\odot$.

\begin{figure}
\caption{\label{fig:Vector} \it \small Plots of contours of constant $T_{\rm kd}$ (left) and $M_{\rm cut}$ (right) for the case of a vector operator interaction between WIMPs and SM particles in which WIMPs couple directly to only leptons or quarks. Shaded regions represent regions of parameter space excluded by collider and direct detection results, while the solid gray curve represents the correct relic density.}
		\subfloat [\it \small As in figure \ref{fig:Vector_univ}, but for DM scattering off leptons only.]{
			\includegraphics[width=.49\textwidth,clip]{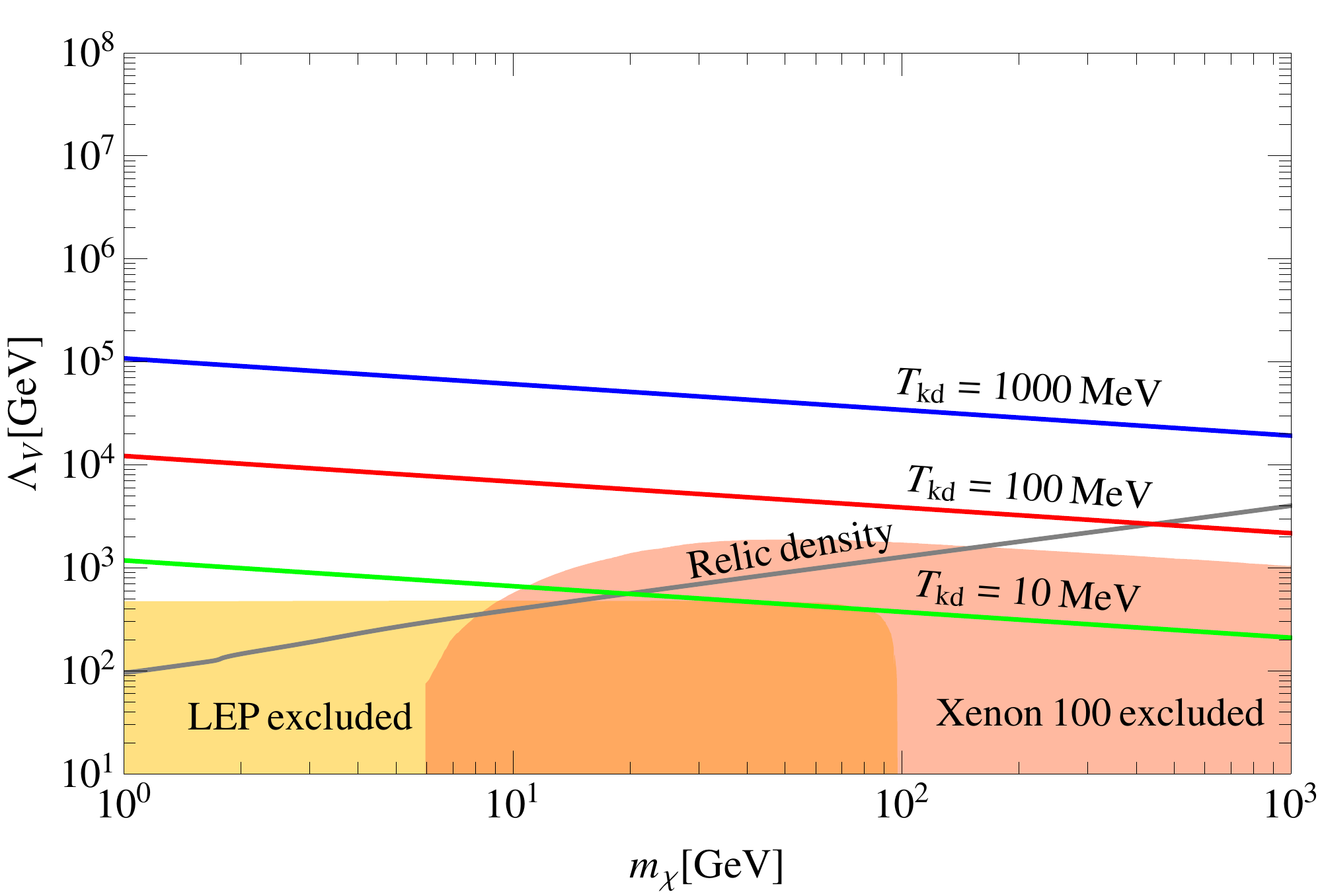}
			\includegraphics[width=.49\textwidth,clip]{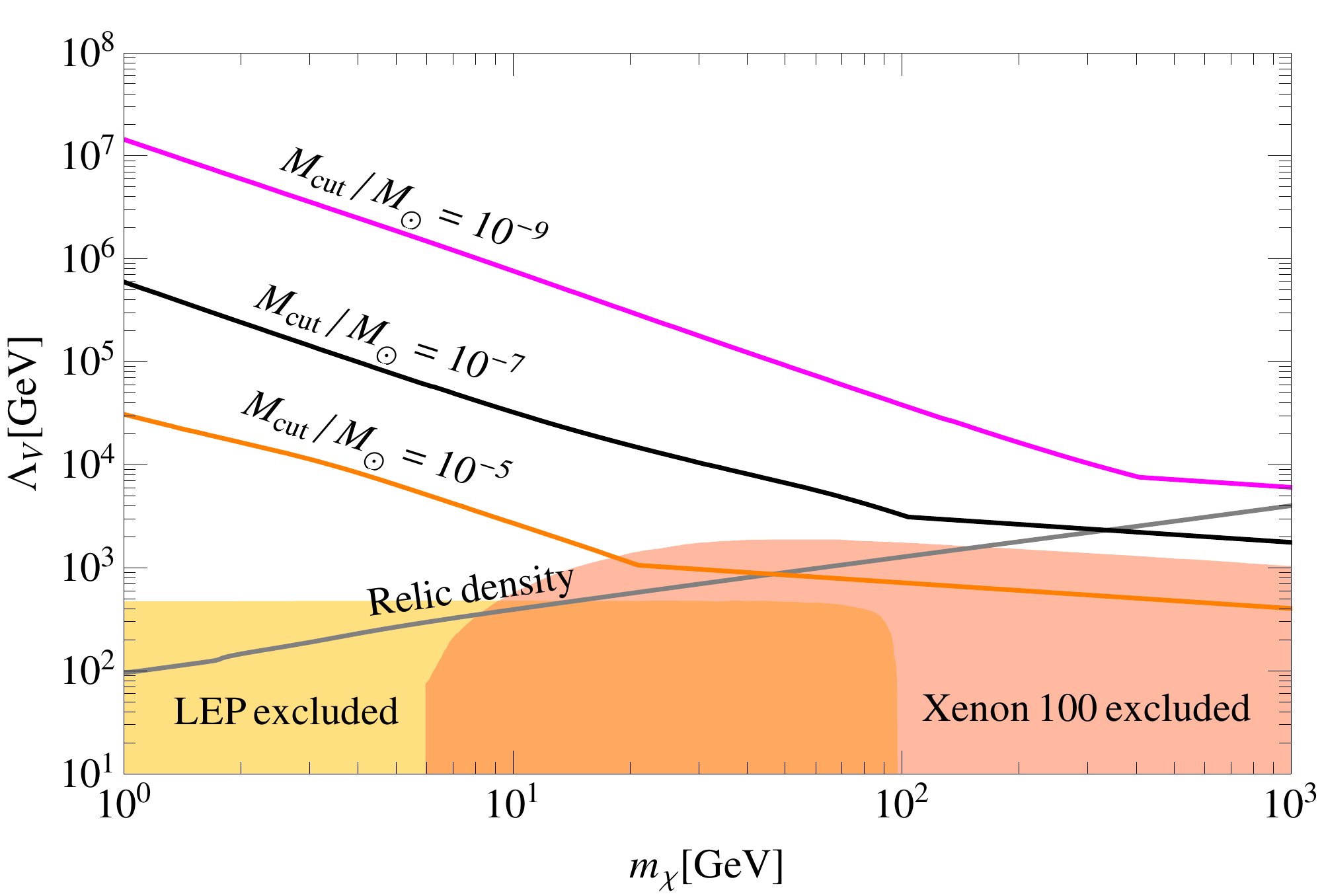}
			\label{fig:Vector_leptons}}\hfill
		\subfloat [\it \small As in figure \ref{fig:Vector_univ}, but for DM coupling directly only to quarks and to leptons via a loop process.]{
			\includegraphics[width=.49\textwidth,clip]{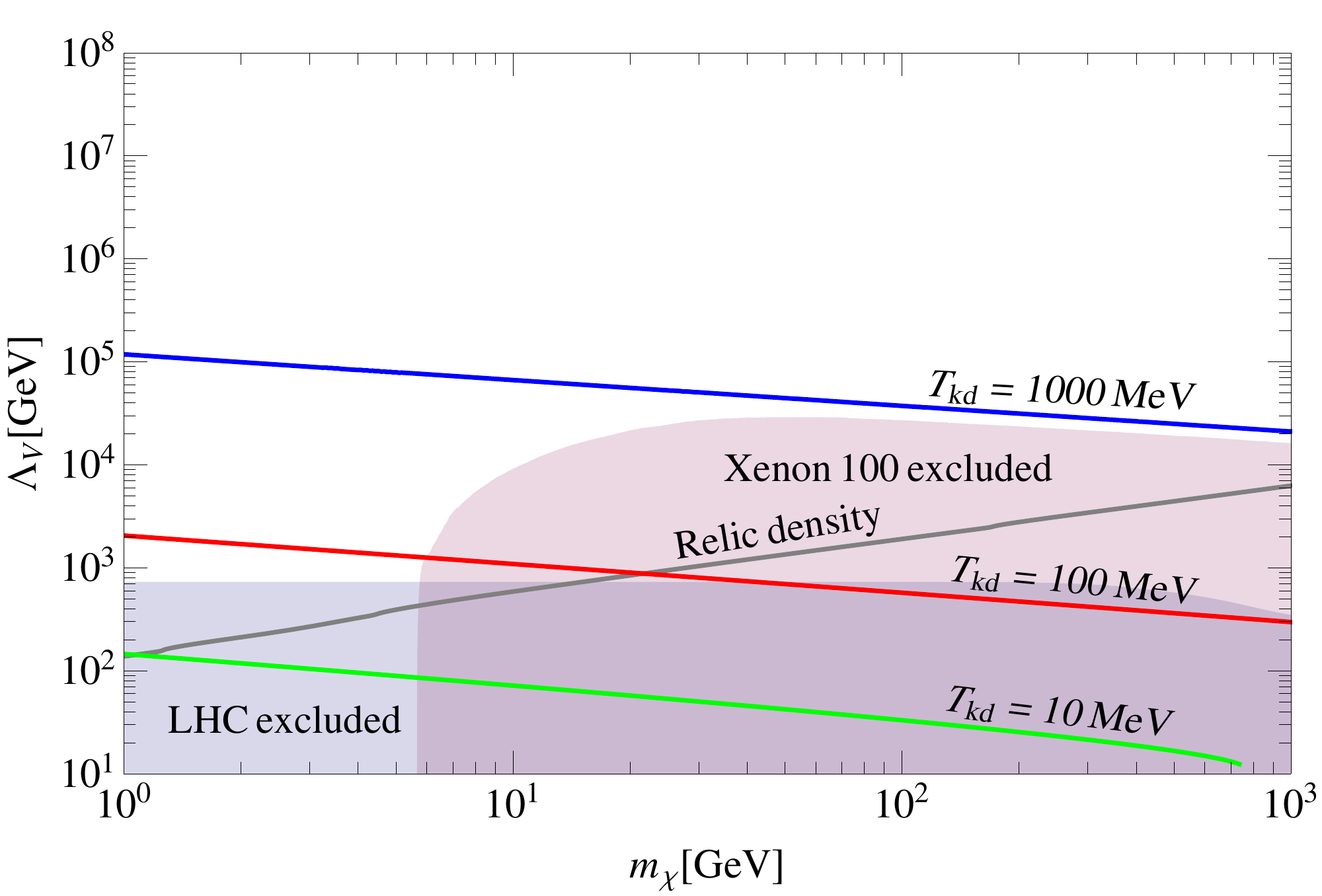}
			\includegraphics[width=.49\textwidth,clip]{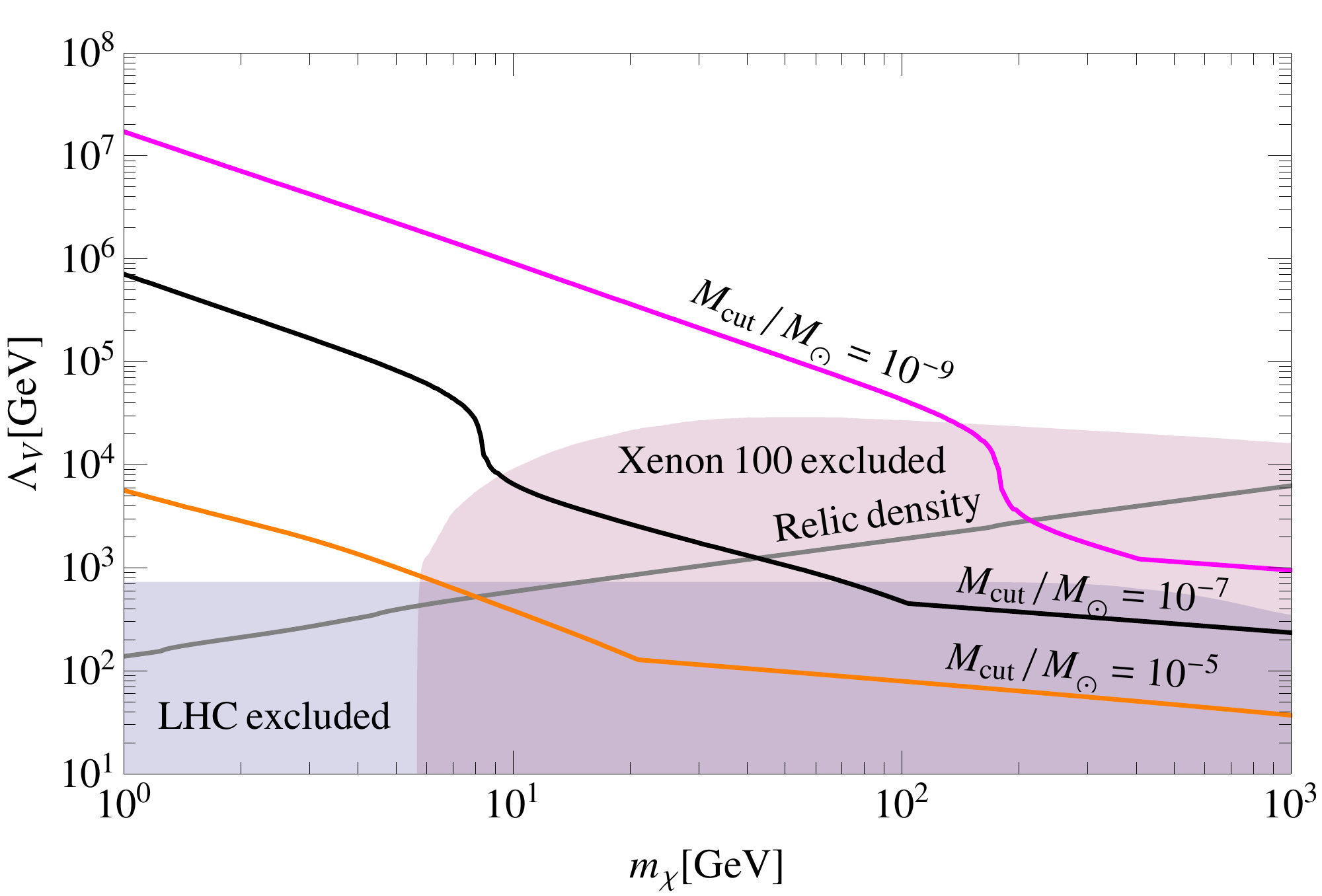} \label{fig:Vector_quarks}}\hfill
		\subfloat [\it \small As in figure \ref{fig:Vector_univ}, but for DM coupling directly to quarks and to either leptons via a loop process (dotted lines) or to pions (dashed lines).]{
			\includegraphics[width=.49\textwidth,clip]{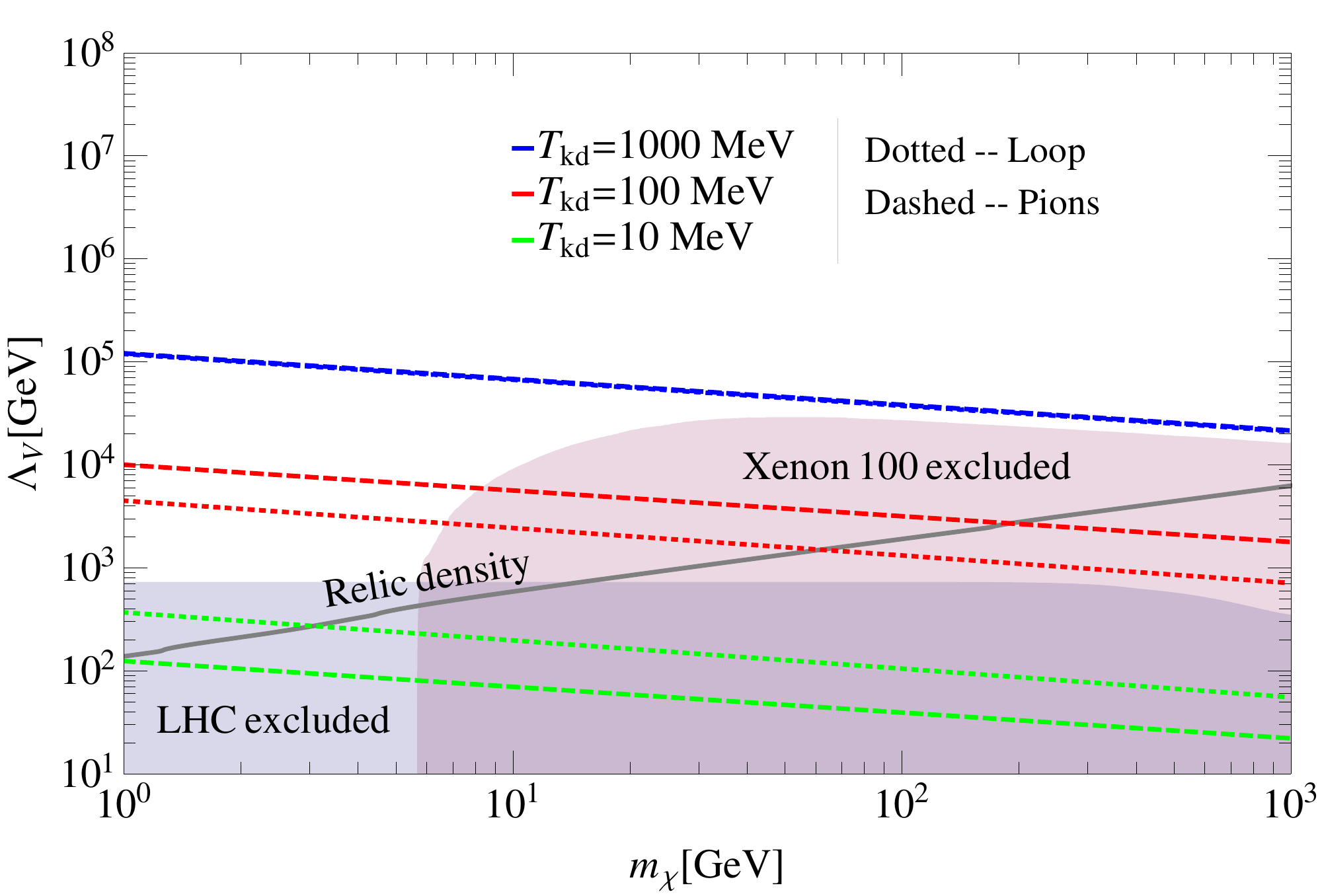}
			\includegraphics[width=.49\textwidth,clip]{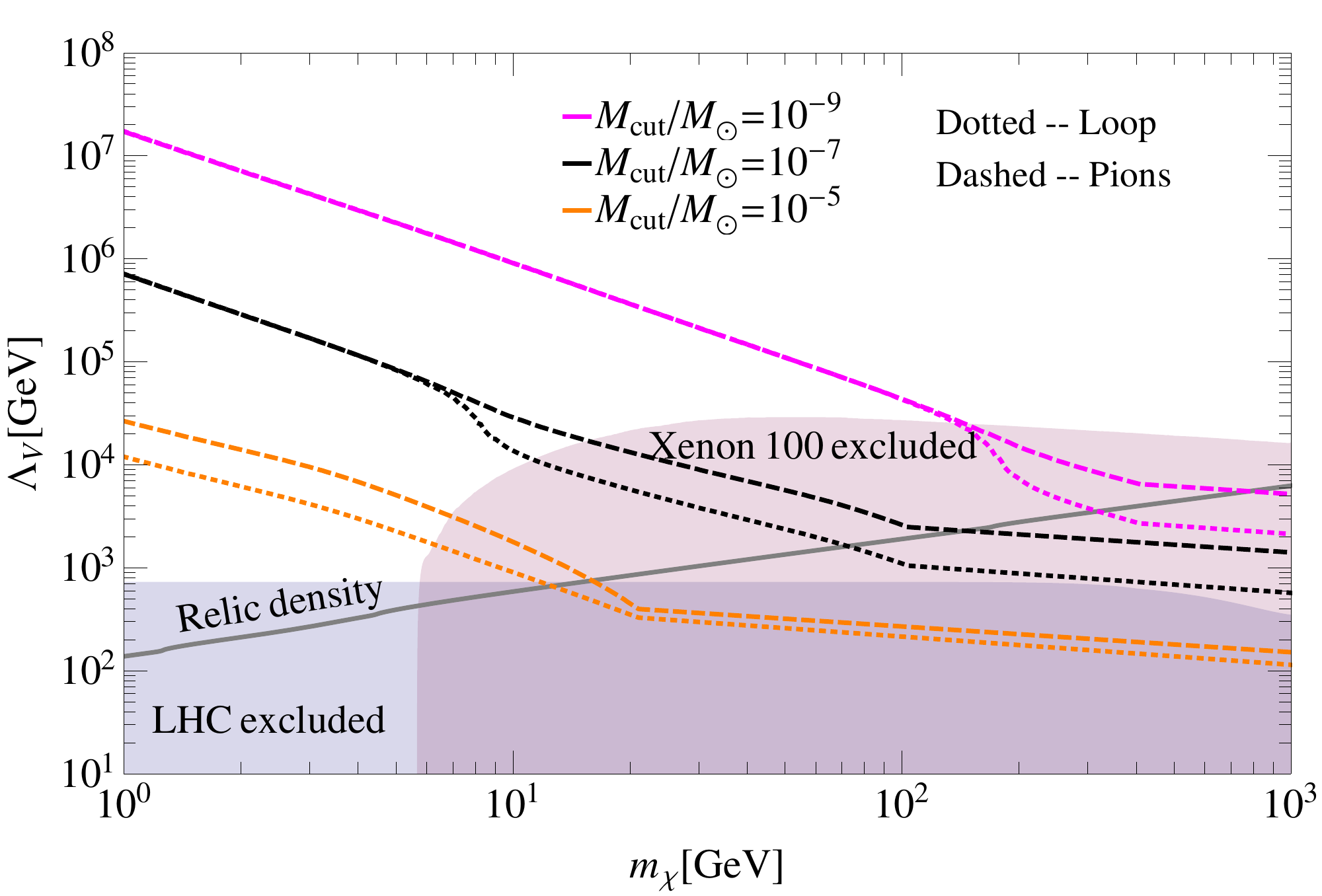}\label{fig:Vector_quarks2}}\hfill
\end{figure}

Our results for vector interactions only with leptons are shown in figure \ref{fig:Vector_leptons}. We have plotted bounds from LEP and from direct detection, which arise at one-loop level by effectively inducing a mixing between the integrated-out heavy vector boson and the SM photon. This mechanism was first discussed by Fox et. al. \cite{Fox:2011fx}, and the bounds we plot are updates of those they derived from the first release of Xenon 100 data to take in to account the full 2012 data set. Even with a loop suppression, direct detection is still the dominant bound on dark matter models with masses above about 8 GeV, and the decoupling temperature is required to be of order 100 MeV. The resulting smallest possible protohalos are smaller than about $10^{-5}\ M_\odot$ for WIMP masses below 100 GeV, and are generically of order $10^{-7}\ M_\odot$ or smaller for masses above 100 GeV.

Considering couplings to quarks only below the QCD phase transition, we now must evaluate $\langle\pi^a|\bar q\gamma^\mu q|\pi^a\rangle$. This also was shown in \cite{kamionkowskietal}, using the conservation of the vector current, to be
\begin{equation}
\langle\pi^a|\bar q\gamma^\mu q|\pi^a\rangle=\left(a_u-a_d\right)\langle\pi^a|\vec\pi\times\partial^\mu\vec\pi|\pi^a\rangle,
\end{equation}
where $a_q$ is the coupling to quarks of type $q$. This clearly vanishes for the coupling structure we have chosen of universal couplings to all quark flavors. 

For a vector interaction coupling only to quarks, the induced direct detection amplitude in the former case can be effectively inverted to give an induced, loop-level coupling to leptons, which can be important as leptons are generically greater contributors to kinetic decoupling than quarks. Since kinetic decoupling is dominated by scatterings at low dark matter velocities, the loop induced coupling to leptons from quarks can be considered as a simple rescaling of the suppression scale involved, and can be calculated in the ``leading-log'' approximation as discussed in Ref.~\cite{ZupanKoppetal}. The formula for this rescaling in the case where we consider identical couplings to all quark flavors is
\begin{equation}
\label{eq:loop}
\Lambda_\ell=\sqrt{\frac{3\pi}{2\alpha}}\frac{\Lambda_q}{\sqrt{\sum_{d,s,b}\ln\left(m_q/\Lambda_q\right)-2\sum_{u,c,t}\ln\left(m_q/\Lambda_q\right)}},
\end{equation}
where $\alpha$ is the electromagnetic fine structure constant and all running quantities are evaluated at the renormalization scale of $\Lambda_q$, the scale of the effective operator. While this does not minimize the logarithms involved, it does allow us to neglect renormalization running and mixing of different operators, such that it is well-defined to assume one operator is dominant. We enforce perturbativity of this loop expansion by truncating our results when the induced coupling to leptons is not weaker than the initial coupling to quarks. The results for this coupling structure are presented in figure \ref{fig:Vector_quarks}. We find, as in fig.~\ref{fig:Vector_univ}, that no models with sub-TeV masses have the right thermal relic density, and that the predicted cutoff for masses above 10 GeV is always smaller than $10^{-7}\ M_\odot$, while it is smaller than $10^{-9}\ M_\odot$ for masses above 100 GeV.

To explore the relative importance of pion scattering to that of loop-induced lepton scattering, we also consider a quarks-only vector-like operator which couples with opposite sign to up- and down-type quarks. This doesn't change the bounds from colliders or the scattering amplitudes above the QCD phase transition, but it allows for pion scattering below the QCD phase transition and alters the bounds from direct detection and the loop-induced coupling to leptons by changing the relative sign of the up- and down-type contributions in Eq.~(\ref{eq:loop}). We have presented the results for this coupling structure in figure \ref{fig:Vector_quarks2}. This plot only shows results for including the coupling to pions or the loop coupling to leptons, but including both contributions leads to a curve that lies along the curve of larger suppression scale: e.g. for $T_{\rm kd} = 100 \, {\rm MeV}$ the curve with both effects included would lie along the pion only curve. From this plot, we observe that following the QCD phase transition, pion scattering is the dominant process in setting $T_{\rm kd}$, but as $T$ decreases, the loop coupling to leptons becomes more important as the pions become non-relativistic and their interaction rate with dark matter is Boltzmann suppressed. We thus note that the relative importance of scattering off of pions versus loop-mediated scattering off of leptons below the QCD phase transition is generically comparable, with one process dominating over the other depending upon the kinetic decoupling temperature: for large decoupling temperatures, hence closer to the QCD confinement phase transition, scattering off of pions dominates, while lepton loop-induced scattering dominates as the number density of pions declines at lower temperatures.

\subsection{Pseudoscalar Operator}

Pseudoscalar operators present a unique complication among all parity-conserving operators, in that the scattering amplitude vanishes in the limit $t\to0$ even when the center-of-mass velocity is large. This necessitates a summation over angles which can be neglected in the case of the other operators, as described in section \ref{sec:tkd}.

\begin{figure}
\caption{\label{fig:Pseudoscalar} \it \small Plots of contours of constant $T_{\rm kd}$ (left) and $M_{\rm cut}$ (right) for the case of a pseudoscalar operator interaction between WIMPs and SM particles. Shaded regions and the dashed curve represent regions of parameter space excluded by collider results, while the solid gray curve represents the correct relic density.}
		\subfloat [\it \small DM couples to all SM fermions.]{\label{fig:PS_univ}
			\includegraphics[width=.49\textwidth,clip]{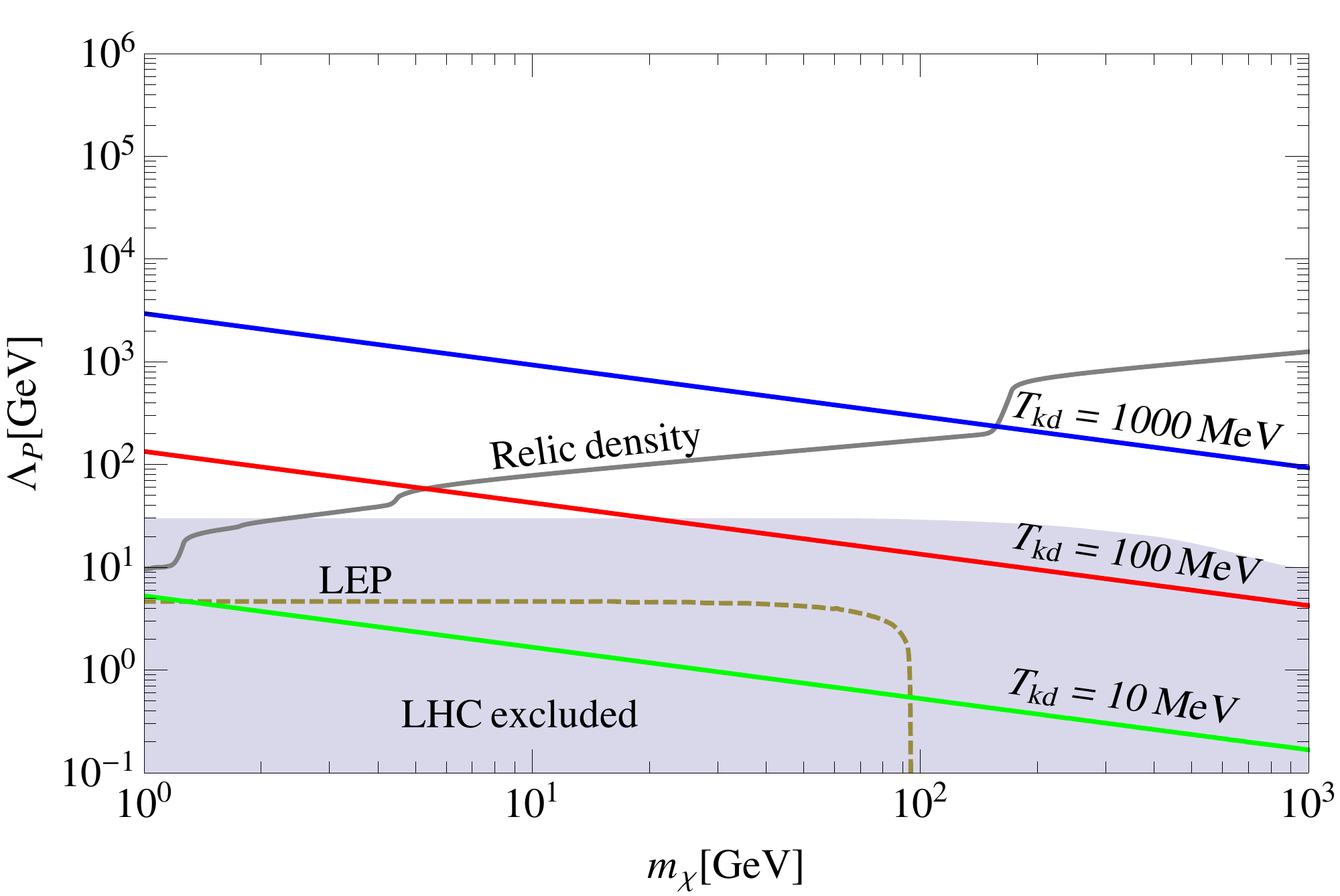}
			\includegraphics[width=.49\textwidth,clip]{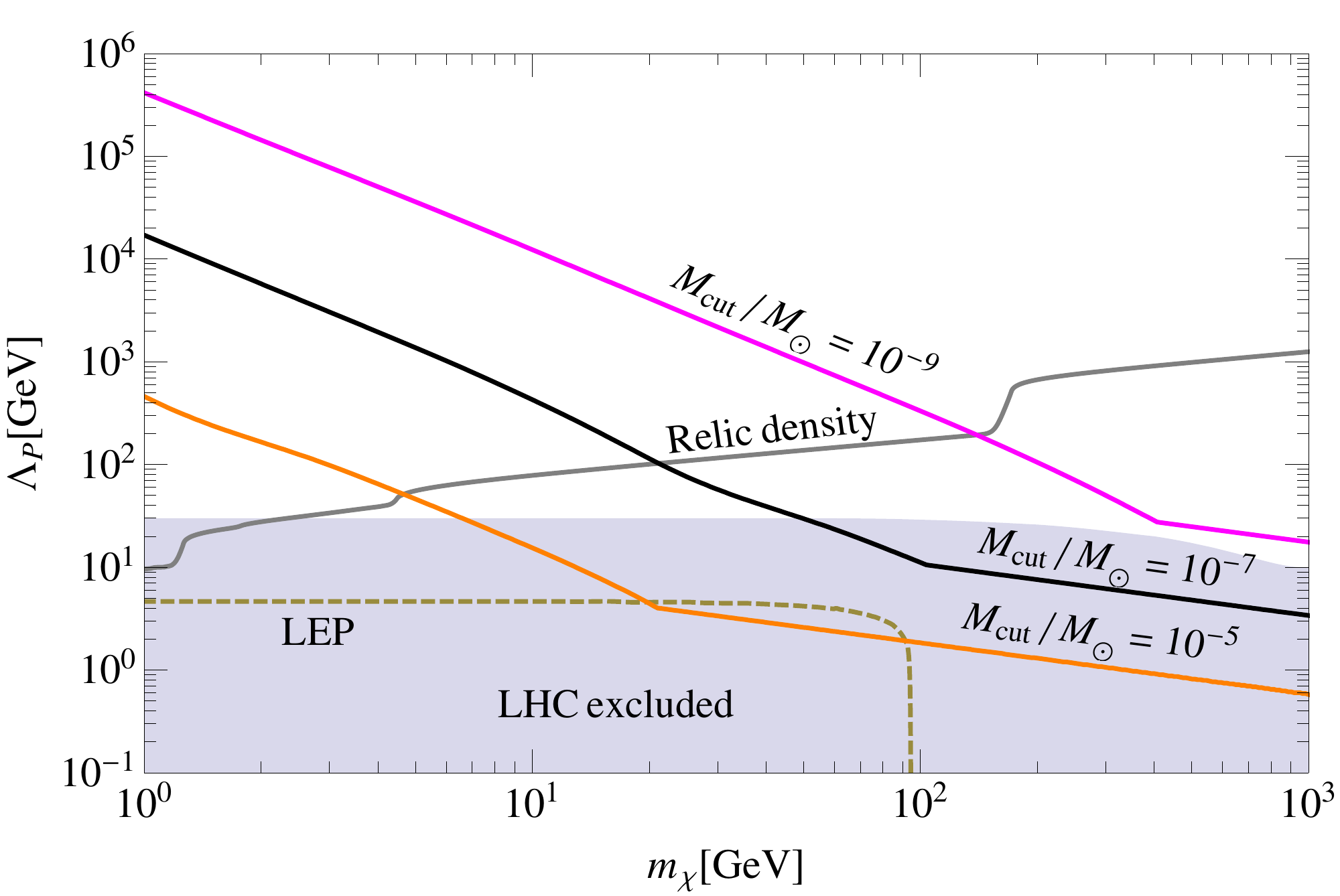}
			}\hfill
		\subfloat [\it \small DM couples only to leptons.]{\label{fig:PS_leptons}
			\includegraphics[width=.49\textwidth,clip]{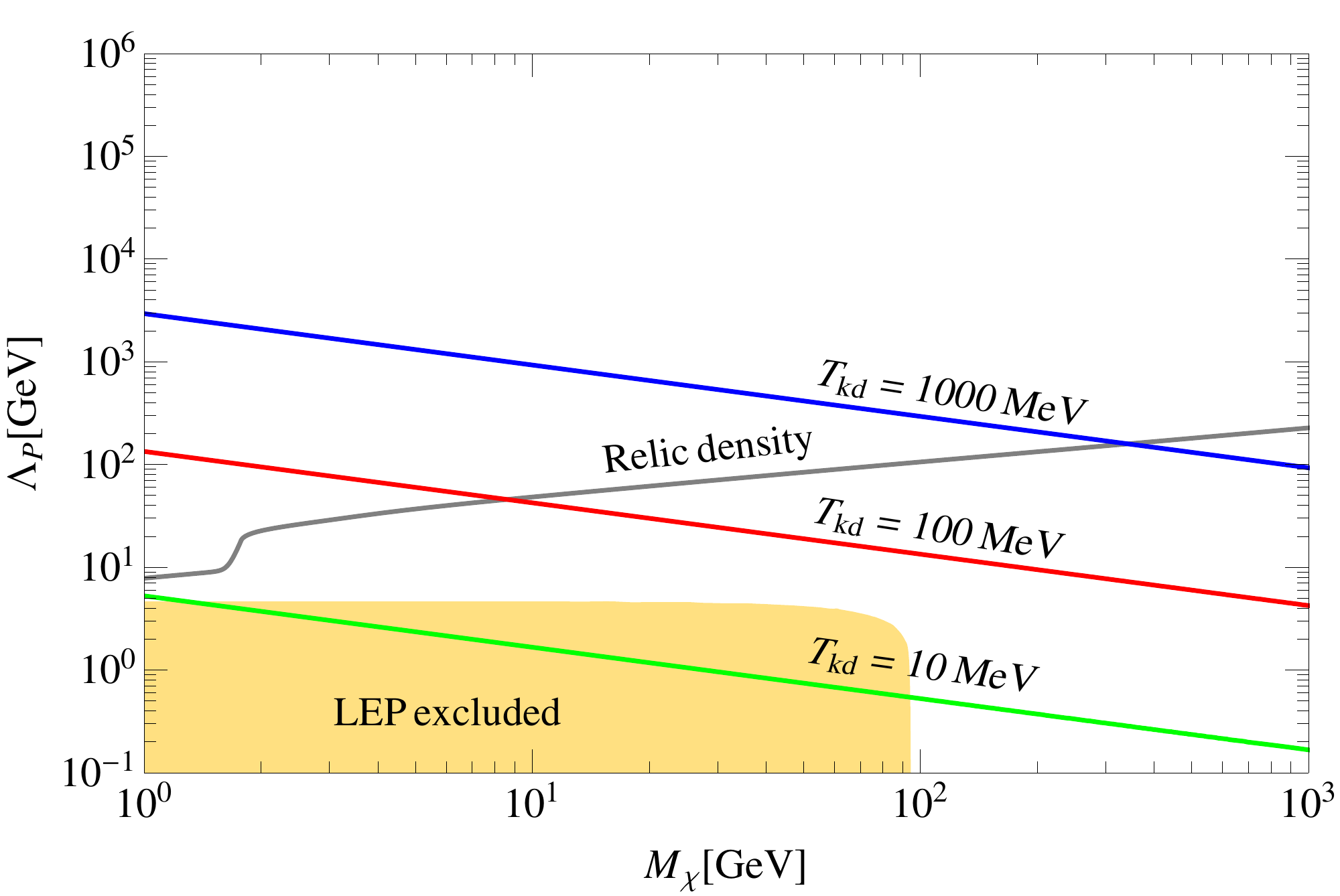}
			\includegraphics[width=.49\textwidth,clip]{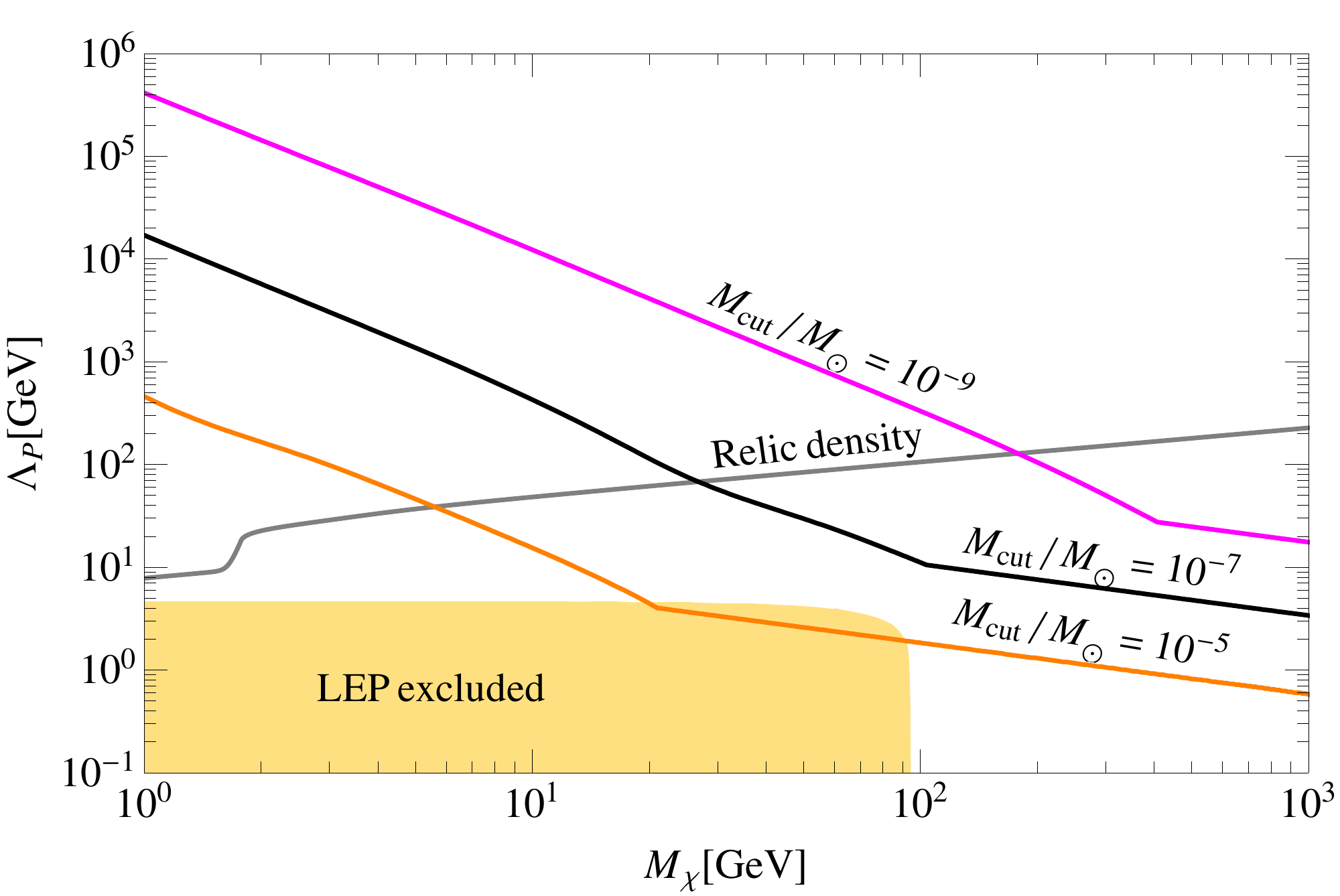}}\hfill
\end{figure}

Pseudoscalar operators lead to strongly suppressed direct detection scattering, so the only relevant bounds are from collider searches. Here, when the coupling is universal, the largest possible value for $M_{\rm cut}$ is $10^{-6} M_\odot$ when $m_\chi \geq 20 \, {\rm MeV}$, as shown in figure \ref{fig:PS_univ}. Models with the correct relic density have masses of a few GeV and higher and increasingly suppressed cutoff scales as a function of mass: from $10^{-5} M_\odot$ at 5 GeV to $10^{-7} M_\odot$ at 20 GeV, and downward to $10^{-9} M_\odot$ and smaller for any mass larger than 200 GeV.

With lepton only couplings constrained by just LEP data, $M_{\rm cut}$ is again much less constrained, as the next figure, \ref{fig:PS_leptons}, shows. Focusing again on models with the correct relic density, we find kinetic decoupling temperatures from slightly more than 10 MeV at WIMP masses in the GeV range, up to 1 GeV for 400 GeV WIMPs. The inferred cutoff mass scale ranges from $10^{-5} M_\odot$ at 6 GeV to $10^{-7} M_\odot$ at 30 GeV, to $10^{-9} M_\odot$ and smaller for any mass larger than 200 GeV, again for models with the correct thermal relic density.

For quark-only pseudoscalar couplings there exists a minimum value of $T_{\rm kd}$, irrespective of how strongly the dark matter couples, which is the QCD phase transition temperature. After the phase transition the only hadronic state available with cosmologically-relevant number densities are the pions. Since QCD is a parity-conserving theory, we can require that the parity behavior of the quark bilinear match that of the pion state which the dark matter would couple to. However, with only two pions it is impossible to construct any pseudoscalar invariant. This indicates that elastic scattering off of two pions is completely forbidden by the symmetries of the theory for this operator. Other scattering processes are possible, however. Inelastic scatterings, whether changing the number of particles or changing, for example, a pion into a sigma meson, are allowed by the symmetries of the problem. These nonetheless do not contribute efficiently to the continued thermalization of the dark matter kinematics, because the thermal bath is not energetic enough to produce the more exotic (i.e. higher-mass) QCD states or to provide sufficient energy to produce additional pions in scattering. Thus, the leading contribution is a one-loop-suppressed process requiring two insertions of the operator, which is very strongly suppressed.

\subsection{Axial-Vector Operator}

\begin{figure}
\caption{\label{fig:Axial} \it \small Plots of contours of constant $T_{\rm kd}$ (left) and $M_{\rm cut}$ (right) for the case of a axial-vector operator interaction between WIMPs and SM particles. Shaded regions represent regions represent regions excluded by collider results and the non-solid curves represent regions of parameter space excluded by collider and direct detection results. The solid gray curve curve represents the correct relic density.}
		\subfloat [\it \small DM couples to all SM fermions.]{\label{fig:Axial_univ}
			\includegraphics[width=.49\textwidth,clip]{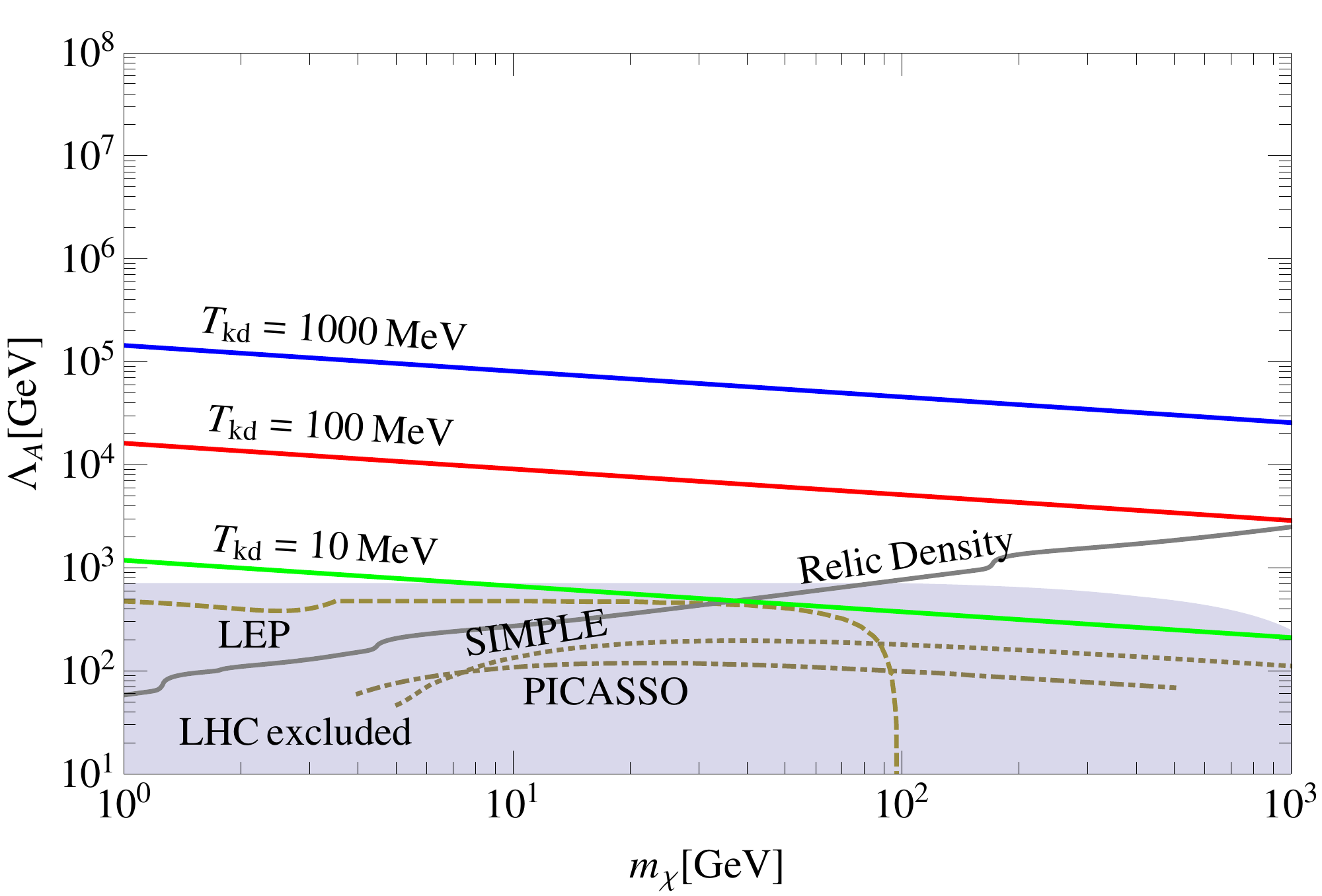}
			\includegraphics[width=.49\textwidth,clip]{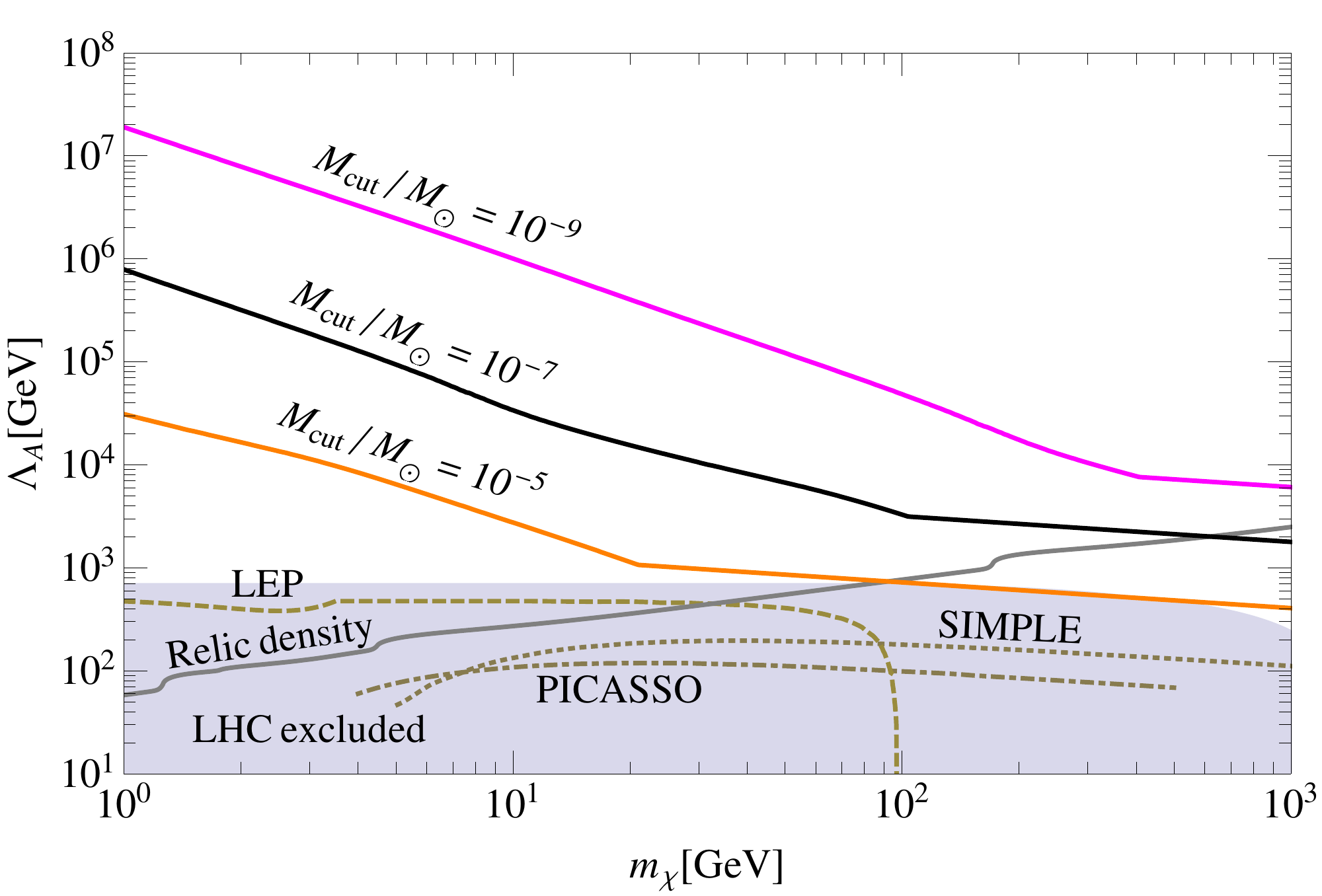}
			}\hfill
		\subfloat [\it \small DM couples only to leptons.]{\label{fig:Axial_leptons}
			\includegraphics[width=.49\textwidth,clip]{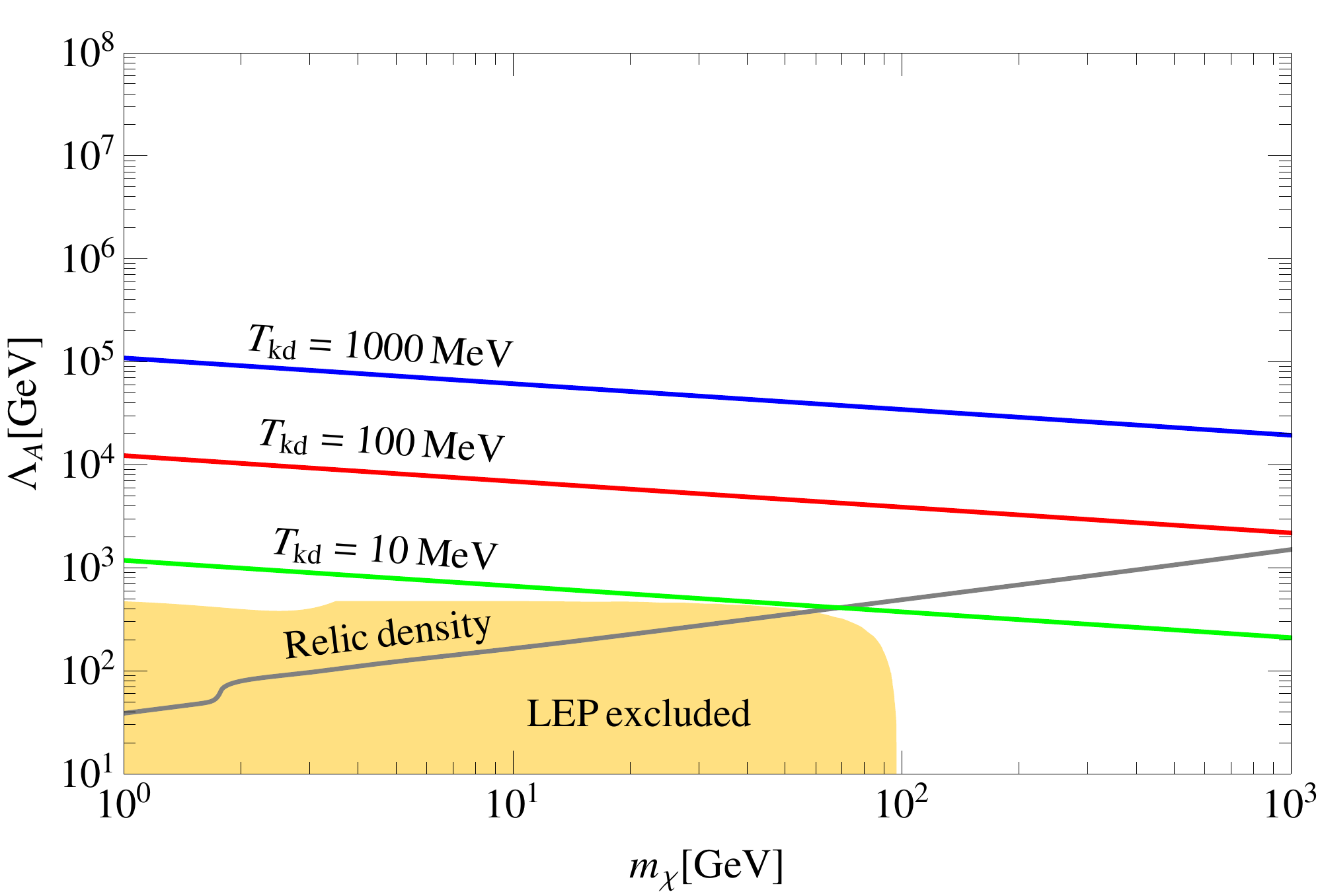}
			\includegraphics[width=.49\textwidth,clip]{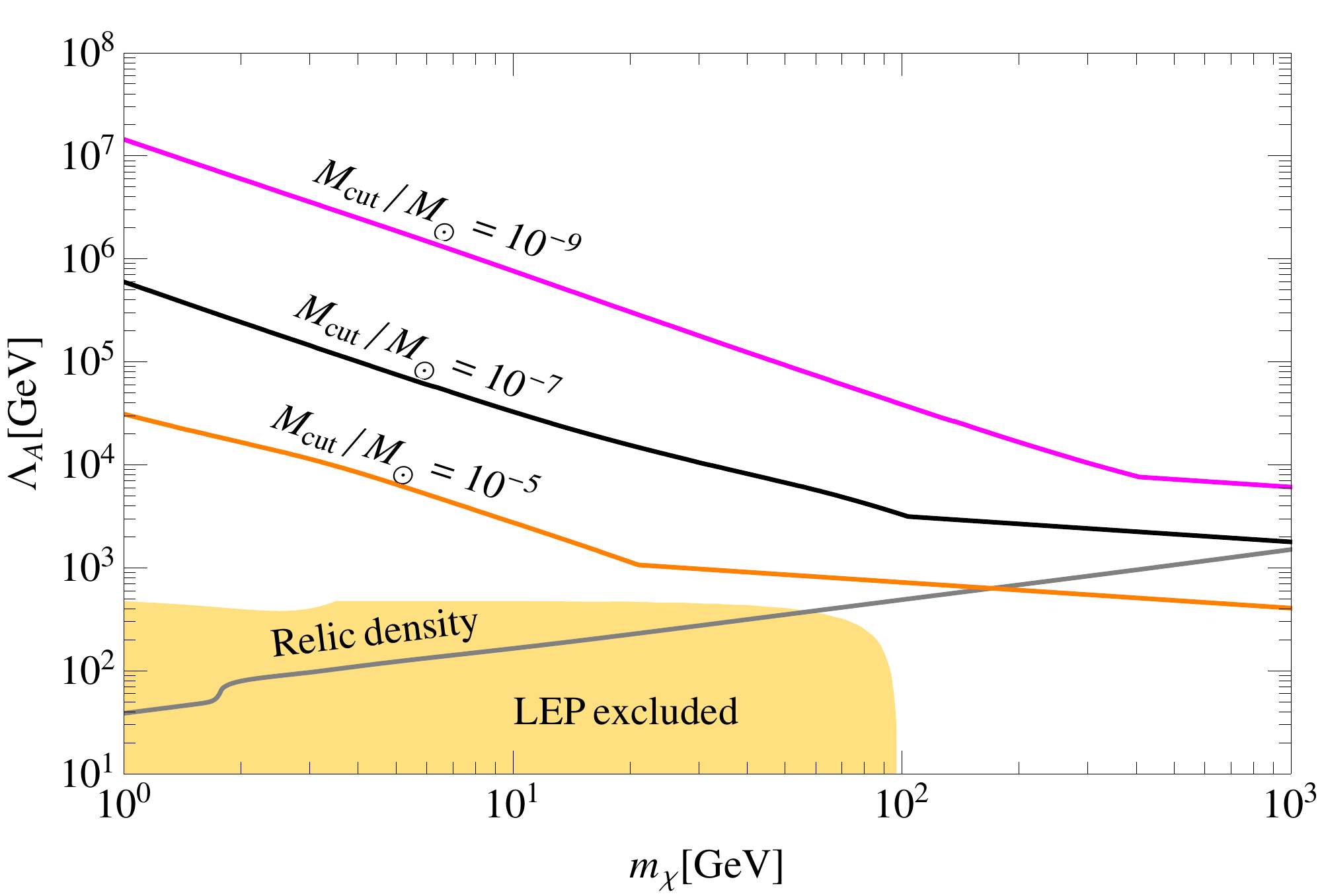}}\hfill
\end{figure}

Axial-vector couplings are constrained at levels comparable to vector couplings by colliders, but lead to spin-dependent rather than -independent scattering in direct detection, so the collider bounds are generically stronger than the direct detection bounds for these interactions. Universal couplings to SM fermions are presented in figure \ref{fig:Axial_univ}. We find that with universal couplings existing bounds generically require $T_{\rm kd}$ to be above 10 MeV, and the cutoff is smaller than $10^{-5}\ M_\odot$ for any mass above 20 GeV.  This class of operators produces the right thermal relic density for WIMPs above 100 GeV, leading in all cases to cutoff masses smaller than $10^{-5}\ M_\odot$

For couplings to leptons only there are no appreciable direct detection bounds, as any loop-induced scattering akin to that in the vector case would have to proceed through $Z$-boson exchange, and the additional suppression of $t/M_Z^2$ makes such contributions negligible. Thus, only LEP bounds are shown along with our results in figure \ref{fig:Axial_leptons}. The figure indicates that cutoff scales as small as about $10^{-3}\ M_\odot$ are in principle possible for very light WIMPs. The thermal relic density and LEP bounds put the dark matter mass in the 100 GeV and up range, with cutoff scales at most of $10^{-4}\ M_\odot$, as before suppressed with increasing WIMP mass.

For the same reason that there are no bounds from direct detection on leptonic axial-vector couplings, there is no induced lepton coupling in the case of a quark-only interaction. Additionally, elastic scattering of dark matter off of pions vanishes in this model, as there is no axial invariant which can be constructed from the kinematics of two pions. Once again, inelastic scattering, whether producing or destroying an additional pion or scattering a pion into a different QCD state, is possible, but the low temperature below the QCD phase transition makes these possibilities contribute negligibly to the kinetic decoupling. Thus axial interactions with quarks only also have a minimum $T_{\rm kd}=T_c$, analogously with the case of pseudoscalar couplings.

\subsection{Tensor Operator}

\begin{figure}
\caption{\label{fig:Tensor} \it \small Plots of contours of constant $T_{\rm kd}$ (left) and $M_{\rm cut}$ (right) for the case of a tensor operator interaction between WIMPs and SM particles, where the solid gray curve represents the correct relic density.}

		\subfloat [\it \small DM couples to all SM fermions.]{\label{fig:Tensor_univ}
			\includegraphics[width=.49\textwidth,clip]{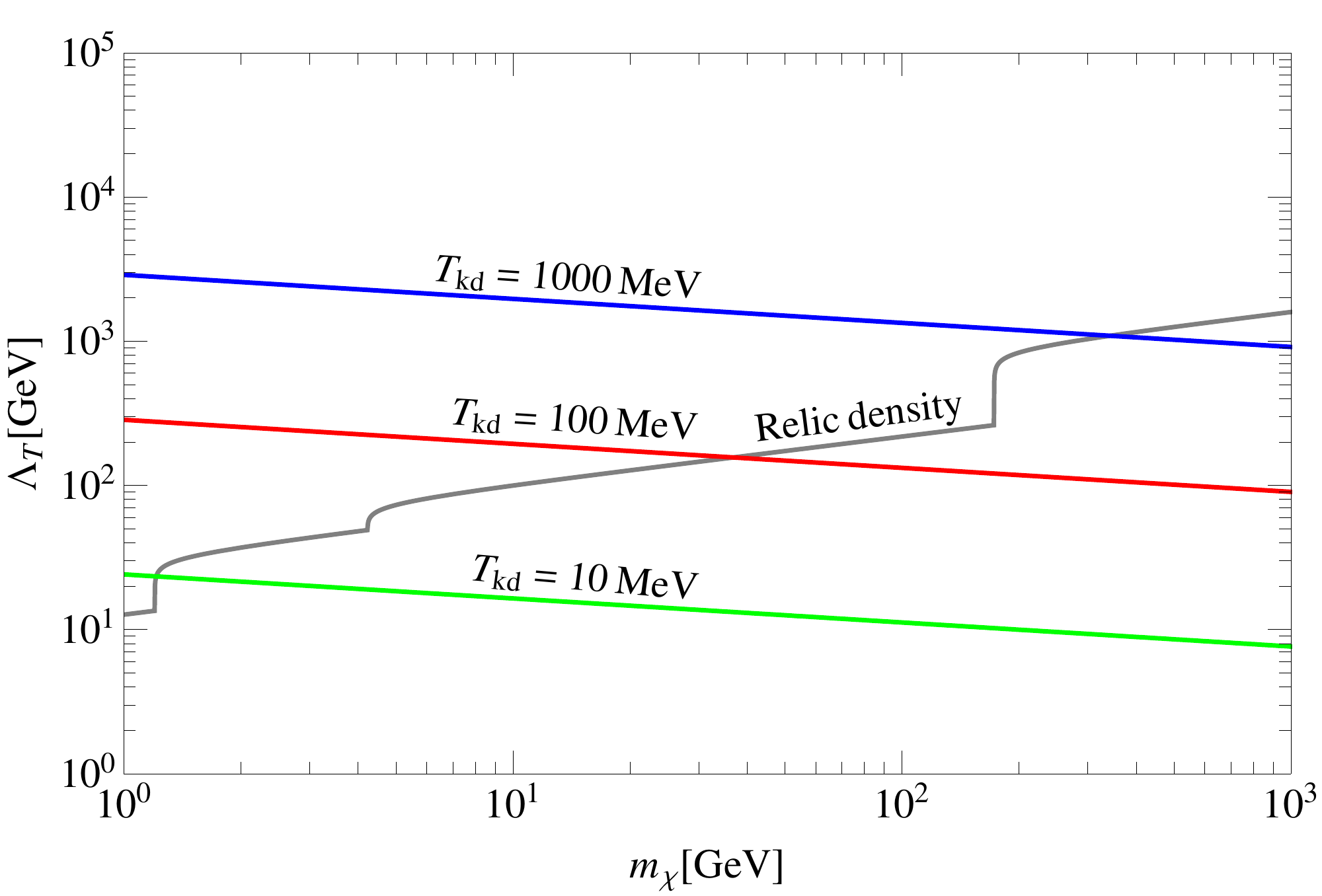}
			\includegraphics[width=.49\textwidth,clip]{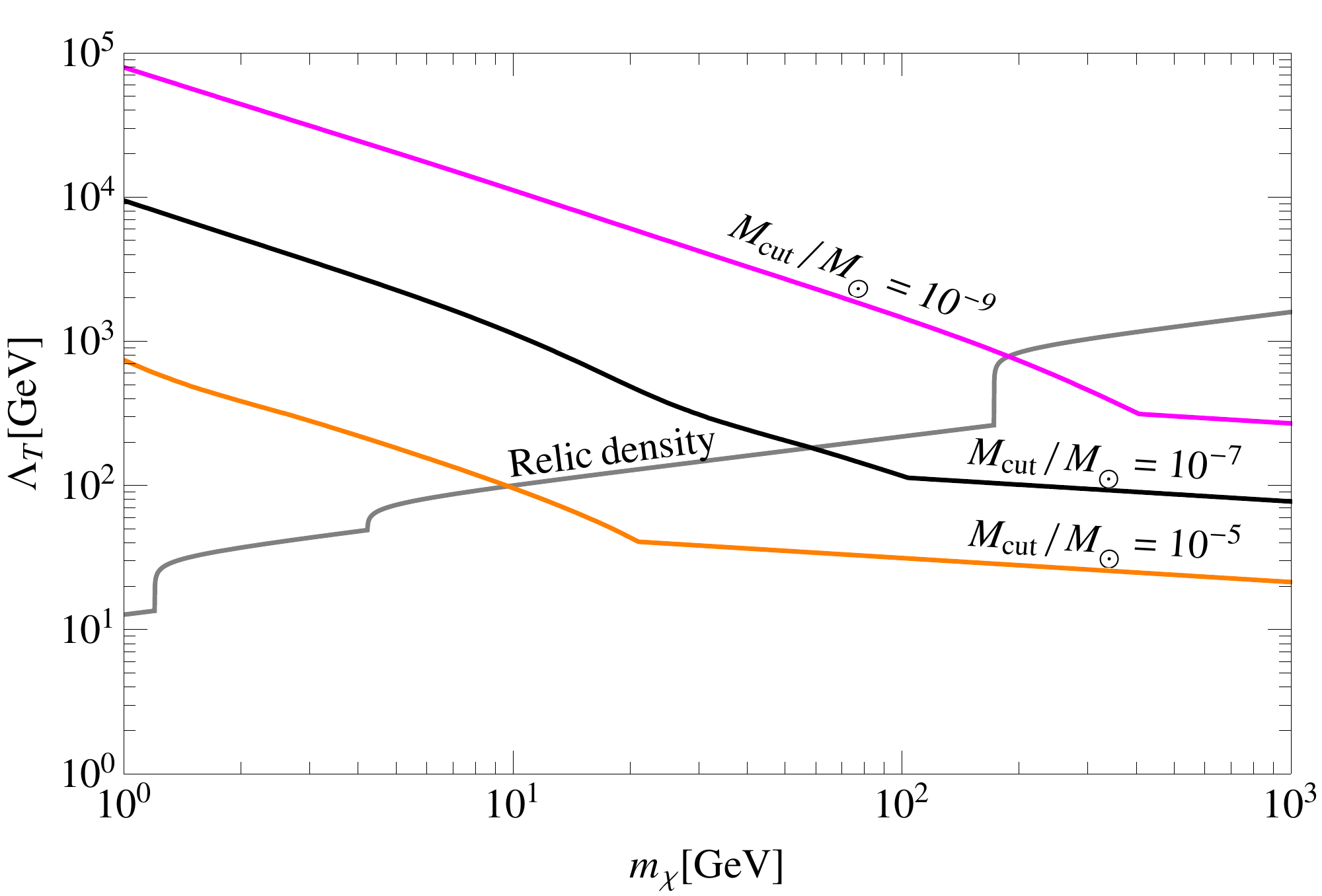}
			}\hfill
		\subfloat [\it \small DM couples only to leptons.]{\label{fig:Tensor_leptons}
			\includegraphics[width=.49\textwidth,clip]{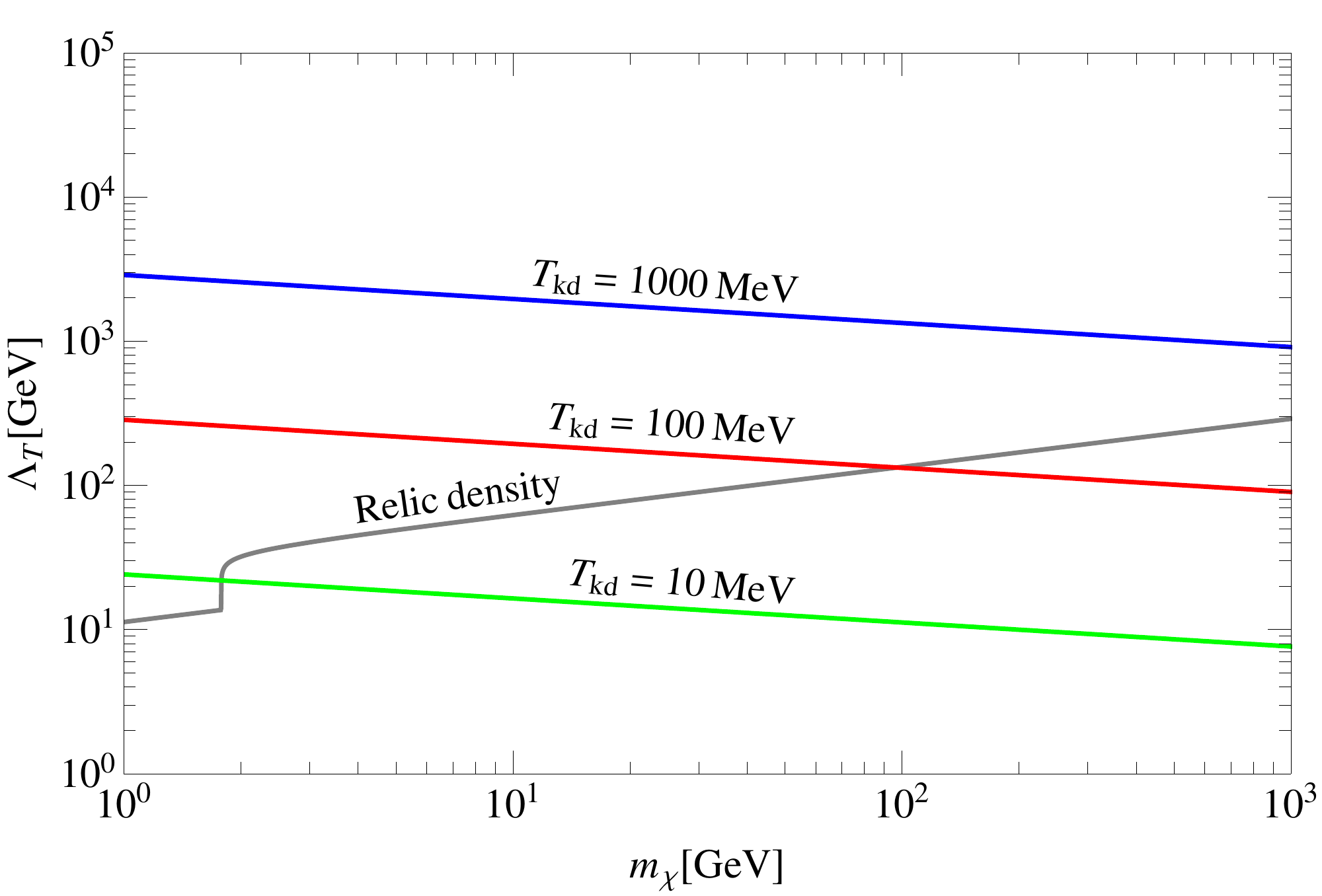}
			\includegraphics[width=.49\textwidth,clip]{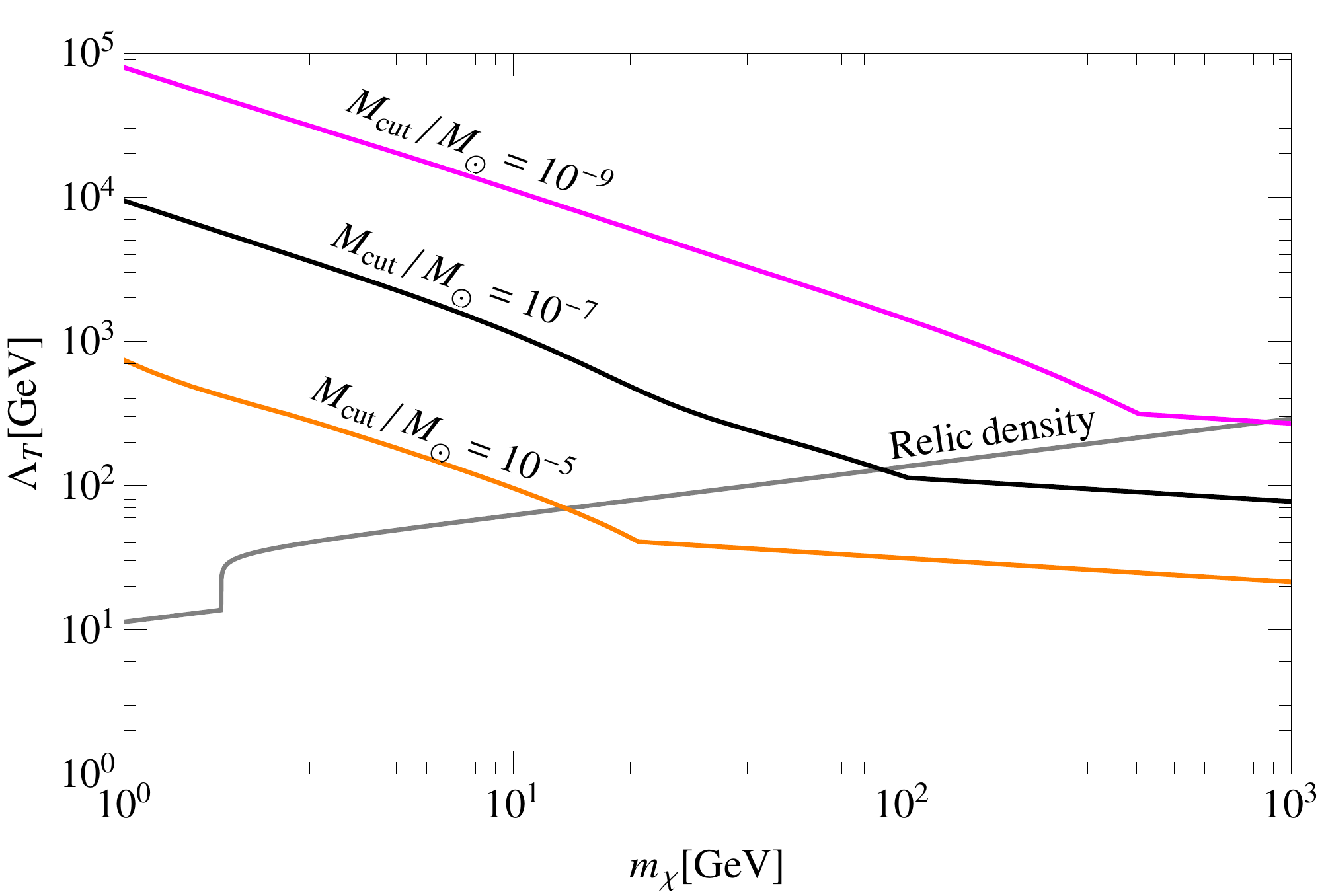}}\hfill
\end{figure}

The tensor operator normalization which we consider preserves the SM gauge group where other normalizations do not, but is not particularly well studied because it does not correspond to the QCD matrix element which is probed in direct detection. Thus, we cannot present bounds from direct detection on this operator. Additionally, current collider searches have been normalized to correspond to direct detection, so we can't compare directly to those results either. The closest approximation to collider searches which could be considered would be the constraints on the other chirality-suppressed operators, in this paper the scalar and pseudoscalar cases. For the direct detection comparison the theoretical picture is a bit more muddled, as the quark mass which appears in this operator should be taken to be related to the yukawa coupling, which will be affected by renormalization running in the strong phase of QCD, and is therefore nontrivial to factor out of the operator and find a meaningful bound. Since neither comparison technique yields a perfect mapping, we will only discuss the early-universe behavior of the operator.

 The results for universal couplings are given in figure \ref{fig:Tensor_univ}, while those for leptons only are in figure \ref{fig:Tensor_leptons}. As a tensor mediated interaction cannot be implemented in CalcHEP, we do not use micrOMEGAs to calculate the relic density, but rather require the velocity averaged cross section to equal the canonical value for s-wave annihilation that gives the correct relic density, i.e
\begin{equation}
\left<\sigma v_{\rm rel} \right> = \sum_f \frac{9}{2 \pi} \frac{m_f^2 m_\chi^2}{\Lambda_T^6} \left(1 - \frac{m_f^2}{m_\chi^2} \right)^{1/2} \approx 3 \times 10^{-26} \, \frac{\rm cm^3}{\rm s} \, ,
\end{equation}
where the sum is over all kinematically accessible fermion annihilation products. Once again, after QCD confines in the early universe there is no pion configuration which has the Lorentz transformation properties of a tensor, and thus quark couplings become irrelevant below $T_c$.

Fig.~\ref{fig:Tensor_univ} indicates that for good thermal relics, the expected kinetic decoupling temperatures are of 10 MeV in the few GeV mass range, up to 100 MeV for a 30 GeV WIMP, and to 1 GeV for a 300 GeV WIMP. The resulting small-scale cutoff masses are of $10^{-5}\ M_\odot$ for a 10 GeV WIMP mass, decreasing to below $10^{-9}\ M_\odot$ for masses above 200 GeV.  For the exclusively leptonic coupling tensor case, we find a qualitatively similar behavior, with good thermal relics producing slightly lower kinetic decoupling temperatures, and slightly larger cutoff scales.

\section{Discussion and Conclusions}\label{sec:concl}

In this study, we addressed the question of establishing the small-scale cutoff of the cosmological matter power spectrum in a variety of particle dark matter models where WIMP coupling to Standard Model fermions is described by effective operators. We included cases where the dark matter separately couples exclusively to leptons, exclusively to quarks, or universally to both leptons and quarks. We also used collider searches and dark matter direct detection to set model-independent limits on the largest experimentally viable value of the small-scale cutoff resulting from kinetic decoupling for each class of operators, and we calculated the dark matter thermal relic abundance on the same parameter space. 

The largest possible cutoffs are found for theories where the dark matter exclusively couples to leptons, as a result of the absence of limits from hadron colliders. For the case of coupling to quarks, in some cases direct dark matter searches squeeze the maximal cutoff scale for protohalos to very small values, in some instances much smaller than the Earth mass. Insisting on setups that produce a thermal relic density in accordance with the observed dark matter density we universally find increasingly suppressed small scale cutoffs with increasing dark matter particle masses.

For theories with quark-only couplings, if the kinetic decoupling falls below the QCD confinement phase transition, two effects exhibit an interesting interplay: scattering off of the lightest available hadronic bound states (pions) and loop-mediated scattering off of leptons, this latter process never having been considered before. We showed that depending on the operator's effective energy scale one or the other effect can dominate the kinetic decoupling process.

While there exist many instances of dark matter models where the effective operator description we adopted here does not apply, the present study has wide applicability to a broad range of WIMP models. In addition, our findings provide a model-independent framework where the relevant range for the small-scale cutoff to the matter power spectrum can effectively be predicted. Finally, this study highlights the complementarity of collider and direct detection of dark matter with questions pertaining to the cosmology of dark matter and the formation of structure in the universe.

\begin{acknowledgments}
\noindent  JMC is supported by the National Science Foundation (NSF) Graduate Research Fellowship under Grant No. (DGE-0809125) and by the Swedish Research Council (VR) and the NSF through the NSF Nordic Research Opportunity. WS and SP are partly supported by the US Department of Energy under contract DE-FG02-04ER41268. 
\end{acknowledgments}

\appendix

\section{Standard Model Fermion Scattering Matrix Elements}\label{sec:fermionm}

\subsection{Scalar}
For the effective operator describing the scalar interaction between a SM fermion $f$ and a Dirac fermion $\chi$
\begin{equation}
\mathcal{O}_S = \frac{m_f}{\Lambda_S^3} \bar\chi\chi\bar ff \, ,
\end{equation}
the matrix element for the scattering process between $f$ and $\chi$ squared and summed over initial spin states and averaged over final spin states is of the form
\begin{equation}
\frac{1}{4} \sum_{\rm Spin \; States} \left| \mathcal{M} \right|^2 = 4 \frac{m_f^2}{\Lambda_S^6} \left( {\bf p} \cdot {\bf p'} + m_{\chi}^2 \right) \left( {\bf k} \cdot {\bf k'} + m_f^2 \right) \, ,
\end{equation}
where ${\bf p}$ and ${\bf p'}$ are the incoming and outgoing 4-momentum of the $\chi$ particle respectively and ${\bf f}$ and ${\bf f'}$ are the same for the SM fermion. Setting $t = 0$, this becomes
\begin{equation}
\frac{1}{4} \sum_{\rm Spin \; States} \left| \mathcal{M} \right| = 16 \frac{m_f^4 m_\chi^2}{\Lambda_S^6} \, .
\end{equation}

\subsection{Psuedoscalar}
We now consider the effective operator describing pseudoscalar interactions,
\begin{equation}
\mathcal{O}_P = \frac{m_f}{\Lambda_P^3} \bar\chi \gamma^5 \chi \bar f  \gamma^5 f \, .
\end{equation}
For a scattering process when the interaction is described by this operator, we find
\begin{equation}
\frac{1}{4} \sum_{\rm Spin \; States} \left| \mathcal{M} \right|^2 = 4 \frac{m_f^2}{\Lambda_P^6} \left(m_\chi^2 - {\bf p} \cdot {\bf p'} \right) \left( m_f^2- {\bf f} \cdot {\bf f'} \right) = \frac{m_f^2}{\Lambda_P^6} t^2 \, .
\end{equation}

\subsection{Vector}
Now considering the operator
\begin{equation}
\mathcal{O}_V = \frac{1}{\Lambda_V^2}\bar\chi\gamma^\mu\chi\bar f\gamma_\mu f \, , \label{eq:apvectop}
\end{equation}
\begin{equation}
\frac{1}{4} \sum_{\rm Spin \; States} \left| \mathcal{M} \right|^2 = \frac{8}{\Lambda_V^4} \left[ \left({\bf p} \cdot {\bf k} \right) \left( {\bf p'} \cdot {\bf k'} \right) + \left( {\bf p} \cdot {\bf k'} \right) \left( {\bf p'} \cdot {\bf k} \right) - \left( {\bf p} \cdot {\bf p'} \right) m_f^2 - \left( {\bf k} \cdot {\bf k'} \right) m_\chi^2 + 2m_\chi^2m_f^2 \right] \, .
\end{equation}
As before, we consider only forward scattering, so $t = 0$. Working in the frame where the dark matter particle is stationary, $s = m_\chi^2+2m_\chi \omega + m_f^2$ and the matrix element becomes:
\begin{equation}
\frac{1}{4} \sum_{\rm Spin \; States}\left| \mathcal{M} \right|^2 = 16 \frac {m_\chi^2}{\Lambda_V^4} \omega^2 \, .
\end{equation}

\subsection{Pseudovector}
The axial vector operator is of the form
\begin{equation}
\mathcal{O}_A = \frac{1}{\Lambda_V^2}\bar\chi\gamma^\mu\gamma^5\chi\bar f\gamma_\mu\gamma^5f \, ,
\end{equation}
\begin{equation}
\frac{1}{4} \sum_{\rm Spin \; States} \left| \mathcal{M} \right|^2 = \frac{8}{\Lambda_A^4} \left[ \left({\bf p} \cdot {\bf k} \right) \left( {\bf p'} \cdot {\bf k'} \right) + \left( {\bf p} \cdot {\bf k'} \right) \left( {\bf p'} \cdot {\bf k} \right) + \left( {\bf p} \cdot {\bf p'} \right) m_f^2 + \left( {\bf k} \cdot {\bf k'} \right) m_\chi^2 + 2m_\chi^2m_f^2 \right] \, .
\end{equation}
Once again, taking the two limits $t = 0$ and  $s = m_\chi^2+2m_\chi \omega + m_f^2$, this becomes
\begin{equation}
\frac{1}{4} \sum_{\rm Spin \; States} \left| \mathcal{M} \right|^2 = 16 \frac{m_\chi^2}{\Lambda_A^4} \left( \omega^2 + 2 m_f^2 \right) \, .
\end{equation}
As $m_f \approx 0$ for relativistic fermions, this is essentially the same as the result for the vector operator.

\subsection{Tensor}
Finally, the tensor operator takes the form
\begin{equation}
\mathcal{O}_T = \frac{m_f}{\Lambda_T^3}\bar\chi\sigma^{\mu\nu}\chi\bar f\sigma_{\mu\nu}f
\end{equation}
where $\sigma^{\mu \nu} = (i / 2) [\gamma^\mu, \gamma^\nu]$.
\begin{equation}
\frac{1}{4} \sum_{\rm Spin \; States} \left| \mathcal{M} \right|^2 =  32 \frac{ m_f^2}{\Lambda_T^6} \left(2 \left({\bf p'} \cdot {\bf k} \right) \left( {\bf p} \cdot {\bf k'} \right) - \left( {\bf p} \cdot {\bf p'} \right) \left( {\bf k} \cdot {\bf k'} \right)  +2 \left( {\bf p} \cdot {\bf k} \right) \left({\bf p'} \cdot {\bf k'} \right) + 3 m_\chi^2 m_f^2  \right) \, ,
\end{equation}
and then when $t = 0$ and  $s = m_\chi^2+2m_\chi \omega + m_f^2$, this becomes
\begin{equation}
\frac{1}{4} \sum_{\rm Spin \; States} \left| \mathcal{M} \right|^2 = 64 \frac{m_f^2 m_\chi^2}{\Lambda_T^6} \left(2 \omega^2 + m_f^2 \right) \, .
\end{equation}

\section{Pion Scattering Matrix Elements}\label{sec:pionm}

\subsection{Scalar}
For the scalar pion coupling we have a Lagrangian term \cite{kamionkowskietal}
\begin{equation}
\mathcal{L}\supset \frac{m_\pi^2}{2\Lambda_S^3}\bar\chi\chi\vec\pi\cdot\vec\pi \, ,
\end{equation}
where
\begin{equation}
\vec{\pi} =
\left( \begin{array}{c}
\frac{1}{\sqrt{2}} \left(\pi^+ + \pi^- \right) \\
\frac{i}{\sqrt{2}} \left(\pi^+ - \pi^- \right) \\
\pi^0 \\
\end{array} \right) \, . \label{eq:pionvec}
\end{equation}
Simplifying the dot product, this gives
\begin{equation}
\mathcal{L}\supset \frac{m_\pi^2}{2\Lambda^3}\bar\chi\chi\left(\pi^0\pi^0+2\pi^+\pi^-\right) \, .
\end{equation}
Note that this leads to a Feynman rule which is identical for all pion charges, and the scattering amplitude which we calculate is
\begin{equation}
i \mathcal{M} = \frac{m_\pi^2}{\Lambda_S^3}\bar\chi\chi \, .
\end{equation}
Squaring and averaging over initial spins, then choosing the zero relative velocity limit, gives the final result \begin{equation}
\frac{1}{2} \sum_{\rm Spin \; States} \left| \mathcal{M} \right|^2=\frac{4m_\pi^4m_\chi^2}{\Lambda_S^6} \, .
\end{equation}

\subsection{Vector}
The coupling from the vector operator has the form \cite{kamionkowskietal}
\begin{equation}
\mathcal{L}\supset \frac{2i}{\Lambda_V^2}\bar\chi\gamma_\mu\chi\left(\vec\pi\times\partial^\mu\vec\pi\right)_3
\end{equation}
when we introduce a negative sign in front of the operator in Eq.~\ref{eq:apvectop} for down type quark interactions, as otherwise this term is zero. The relevant component of the cross product is $\pi_1\partial^\mu\pi_2-\pi_2\partial^\mu\pi_1$, which can be rewritten in terms of the physical fields to give 
\begin{equation}
\mathcal{L}\supset \frac{2i}{\Lambda_V^2}\bar\chi\gamma_\mu\chi\left(\pi^+\partial^\mu\pi^--\pi^-\partial^\mu\pi^+\right) \, .
\end{equation}
Thus, this operator does not couple to neutral pions, and the scattering amplitude off of a charged pion is equal to
\begin{equation}
i \mathcal{M}=\frac{2}{\Lambda_V^2}\bar\chi\gamma_\mu\chi\left({\bf k}+{\bf k^\prime}\right)^\mu \, .
\end{equation}
Squaring and averaging over incoming spins, we have
\begin{equation}
\frac{1}{2} \sum_{\rm Spin \; States} \left| \mathcal{M} \right|^2 =\frac{8}{\Lambda_V^4}\left( \left({\bf k}+{\bf k'}\right)^2 \left(m_\chi^2-{\bf p} \cdot {\bf p^\prime}\right)+2{\bf p}\cdot\left({\bf k}+{\bf k^\prime} \right){\bf p^\prime} \cdot\left({\bf k}+{\bf k^\prime}\right)\right) \, .
\end{equation}
Simplifying this in terms of Mandelstam variables we find 
\begin{equation}
\frac{1}{2} \sum_{\rm Spin \; States} \left| \mathcal{M} \right|^2=\frac{4}{\Lambda_V^4}\left(4m_\pi^2t-t^2+s^2+u^2-2su\right) \, ,
\end{equation}
and working in the limit where $t \rightarrow 0$, this becomes
\begin{equation}
\frac{1}{2} \sum_{\rm Spin \; States} \left| \mathcal{M} \right|^2=\frac{64m_\chi^2\omega^2}{\Lambda_V^4} \, .
\end{equation}

\end{document}